\documentclass{aa}  

\usepackage{graphicx}
\usepackage{placeins}
\usepackage{txfonts}
\usepackage{lineno}

\begin{document}
\nolinenumbers

  \title{MICONIC: JWST/MIRI MRS reveals a fast ionized gas
    outflow in the central region of
    Centaurus A}

   \titlerunning{MICONIC: Fast ionized gas outflow in Cen~A with MIRI MRS}
   
   \author{A. Alonso Herrero\inst{1}
         \and
         L. Hermosa Muñoz\inst{1}
         \and
         A. Labiano\inst{2}
         \and
         P. Guillard\inst{3, 4}
          \and
         M. Garc\'{\i}a-Mar\'{\i}n\inst{5}
         \and
         D. Dicken\inst{6}
         \and
         S. Garc\'{\i}a-Burillo\inst{7} 
         \and
         L. Pantoni\inst{8}
         \and
         V. Buiten\inst{9}
         \and
         L. Colina\inst{10}
         \and
         T. B\"oker\inst{5}  
         \and
         M. Baes\inst{8}
         \and
         A. Eckart\inst{11, 12}
         \and
         L. Evangelista\inst{3, 4}
         \and
         G. \"Ostlin\inst{13}
         \and
         D. Rouan\inst{14}
         \and
         P. van der Werf\inst{9}
         \and
         F. Walter\inst{15}
         \and
         M. J. Ward\inst{16}
         \and
         G. Wright\inst{6}
         \and
         M. G\"udel\inst{17, 18}
         \and
         Th. Henning\inst{15}
         \and
         P.-O. Lagage\inst{19}}

       \institute{Centro de Astrobiolog\'{\i}a (CAB), CSIC-INTA,
         Camino Bajo del Castillo s/n, E-28692 Villanueva de la Ca\~nada, Madrid,
     Spain\\
     \email{aalonso@cab.inta-csic.es}\label{inst1}
   \and
       {Telespazio UK for the European Space Agency (ESA), ESAC,
Camino Bajo del Castillo s/n, 28692 Villanueva de la Ca\~nada,
Spain}\label{inst2}
\and
{Sorbonne Universit\'e, CNRS, UMR 7095, Institut d’Astrophysique de
  Paris, 98bis bd Arago, 75014 Paris, France}\label{inst3}
\and
{Institut Universitaire de France, Minist\`ere de l’Enseignement Sup\'erieur et de la Recherche, 1 rue Descartes, 75231 Paris Cedex 05,
  France}\label{inst4}
\and
{European Space Agency, c/o Space Telescope Science Institute, 3700 San
  Martin Drive, Baltimore MD 21218, USA}\label{inst5}
\and
{UK Astronomy Technology Centre, Royal Observatory, Blackford Hill
  Edinburgh, EH9 3HJ, Scotland, UK}\label{inst6}
\and
{Observatorio Astron\'omico Nacional (OAN-IGN)-Observatorio de
  Madrid, Alfonso XII, 3, 28014 Madrid, Spain}\label{inst7}
\and
{Sterrenkundig Observatorium, Universiteit Gent, Krijgslaan 281 S9,
  B-9000 Gent, Belgium}\label{inst8} 
\and
{Leiden Observatory, Leiden University, PO Box 9513, 2300 RA Leiden,
  The Netherlands}\label{inst9} 
\and
{Centro de Astrobiolog\'{\i}a (CAB), CSIC-INTA, Ctra. de Ajalvir km 4,
  Torrej\'on de Ardoz, 28850, Madrid, Spain}\label{inst10}
\and
{I. Physikalisches
Institut der Universit\"at zu K\"oln, Z\"ulpicher Str. 77, D-50937 K\"oln, Germany}\label{inst11}
\and
{Max-Planck-Institut f\"ur Radioastronomie (MPIfR), Auf dem H\"ugel 69,
D-53121 Bonn, Germany}\label{inst12} 
\and
{Department of Astronomy, Stockholm University, The Oskar Klein
  Centre, AlbaNova, SE-106 91 Stockholm, Sweden}\label{inst13}
\and
{LIRA, Observatoire de Paris, Universit\'e PSL, Sorbonne Universit\'e,
  Universit\'e Paris Cit\'e, CY Cergy Paris Universit\'e, CNRS, 
92190 Meudon, France}\label{inst14}
\and
{Max Planck Institute for Astronomy, K\"onigstuhl 17, 69117
  Heidelberg, Germany}\label{inst15}
\and
{Centre for Extragalactic Astronomy, Durham University, South Road,
  Durham DH1 3LE, UK}\label{inst16}
\and
{Dept. of Astrophysics, University of Vienna, T\"urkenschanzstr. 17, A-1180 Vienna, Austria}\label{inst17}
\and
{ETH Z\"urich, Institute for Particle Physics and Astrophysics,
  Wolfgang-Pauli-Str. 27, 8093 Z\"urich, Switzerland}\label{inst18}
\and
{Universit\'e Paris-Saclay, Universit\'e Paris Cit\'e, CEA, CNRS, AIM, 91191 Gif-Sur-Yvette, France}\label{inst19}}

   \date{Received 2025; accepted 2025}

  \abstract
{We present a kinematical study of the ionized and molecular gas in
  the central region ($\sim$7--14\arcsec$\sim$100--200\,pc) of the
  nearby radio galaxy Centaurus A
  (Cen~A). We used JWST/MIRI MRS
  $\sim 5-28\,\mu$m observations taken as part of the Mid-Infrared
   Characterization of Nearby Iconic galaxy Centers 
(MICONIC) of the MIRI European
Consortium. The two gas phases present contrasting morphologies
and kinematics in Cen~A. The brightest emission from the ionized gas, traced
with a range of ionization potential (IP) lines analyzed here (from [Fe\,{\sc ii}]
to [Ne\,{\sc vi}]), is extended along the direction of the radio jet.  We also detected
emission from low IP emission lines and H$_2$ transitions in the
galaxy disk region mapped with MRS. Both gas phases present rotational 
motions in the disk but also complex kinematics. The MRS observations
reveal  several ionized gas kinematical features that are consistent with simulation
predictions of a jet-driven bubble and outflow interacting with the galaxy
interstellar medium. These 
include broad components in the nuclear line profiles ($\sigma\sim
600\,{\rm km\,s}^{-1}$ in [Ar\,{\sc ii}] and [Ne\,{\sc iii}]),  high
velocities (reaching approximately +1000, $-$$1400\,{\rm km\,s}^{-1}$)
confined within the nuclear region, velocities of hundreds of
kilometers per second
in several directions in the central 2\arcsec, and enhanced velocity
dispersions perpendicular to the radio jet. Moreover, we find evidence
of shock excitation in the nuclear region of Cen~A
based on mid-infrared line ratios. We compared the ionized gas mass outflow rate
with Cen~A's active galactic nucleus (AGN) luminosity and radio jet
power and demonstrate that both 
mechanisms provide sufficient energy to launch the outflow. The
noncircular motions observed in the mid-infrared H$_2$ lines can be
reproduced with either a  warped rotating disk model or a radial 
component. The latter might be to related to gas streamers detected in
cold molecular gas. Notably, there is no clear indication of a fast nuclear
H$_2$ outflow in Cen A, only a weak blueshifted component in the line 
profiles. This could be due to a relatively low nuclear warm H$_2$
column density and/or the limited geometrical 
coupling of Cen~A's inner radio 
jet with the circumnuclear disk of the galaxy. }


   \maketitle

\section{Introduction}\label{sec:introduction}

Low- and intermediate-power radio jets,  $P_{\rm jet} \sim 10^{43}-10^{44}\,{\rm erg \,
  s}^{-1}$, are now being recognized as playing an important
role in driving ionized 
and molecular outflows in active galactic nuclei
(AGNs). There are many nearby examples of multiphase outflows in radio-quiet
Seyferts and
low luminosity AGNs (LLAGNs) where the radio jet may play an important
role. Evidence for this is provided by 
optical \citep{Cresci2015, Venturi2021, 
  Cazzoli2022, PeraltadeArriba2023, HermosaMunoz2024LINERS},
  infrared \citep{PereiraSantaella2022, Dasyra2024, Zhang2024},  and
  (sub)millimeter \citep[e.g.,][]{GarciaBurillo2014,  Morganti2015, Audibert2019} 
observations. How radio jet driven outflows compare with those
driven by AGN radiation, how significantly they modify the gas
properties in
their host galaxies, and how they affect their evolution via negative and/or
positive feedback are still open questions \citep[see the review by][]{Harrison2024}.

Centaurus A (hereafter Cen~A, also known as NGC~5128) is the nearest
\citep[distance of 3.5\,Mpc,  1\arcsec=17\,pc, ][]{Neumayer2007}  radio galaxy
\citep[see][for a review]{Israel1998}. It has a 
modest AGN bolometric luminosity  $L_{\rm bol}{\rm (AGN)}
\simeq 1-4 \times 10^{43}\,{\rm
  erg\,s}^{-1}$ \citep{Israel1998, Beckmann2011}, and a large-scale
radio jet extending over a few hundred kiloparsecs. On the nuclear 
scales of interest for this work (inner $\sim 200\,$pc), the jet is oriented at
a position angle of PA$_{\rm jet}=51^{\rm o}$ and displays apparent
subluminal motions \citep{Clarke1992, 
  Hardcastle2003}. It appears to be nearly
perpendicular to the disk on scales of the circumnuclear disk detected
in cold molecular gas \citep[projected size of $20\arcsec \times
10\arcsec$, see][]{Espada2017}.  
The present-day power of its radio jet is
estimated to be on the order of 
$P_{\rm jet} \sim 1-2\times 10^{43}\,{\rm erg \,
  s}^{-1}$ \citep[see e.g.,][]{Croston2009, Wykes2013, Neff2015}.

Simulations of low-intermediate power radio
jets similar to Cen~A predict fast outflows in
the hot and warm (ionized) gas, even when the jet is launched   
perpendicular to the disk of the galaxy.
Jet-inflated bubbles can interact with the interstellar
medium (ISM) in the galaxy disk, producing both inflows and outflows
\citep[see e.g.,][]{Mukherjee2016, Mukherjee2018, Talbot2022,
  Meenakshi2022, Borodina2025}. These  studies also demonstrated  that the
jet-driven outflow properties depend strongly on the jet
power, its relative orientation with respect to the galaxy disk,
and surrounding ISM properties.

In this work we analyze spatially resolved kinematics of the ionized
and warm molecular gas in the
central ($\sim 7-14\arcsec$$\sim$100--200\,pc) region of Cen~A. We  used 
mid-infrared (mid-IR) observations 
 obtained with the 
Mid-Infrared Spectrometer \citep[MRS:][]{Wells2015, Argyriou2023}
of the Mid-InfraRed Instrument 
\citep[MIRI:][]{Rieke2015, Wright2015, Wright2023} on board
 the {\it James Webb} Space 
 Telescope \citep[JWST:][]{Gardner2023}. This work is part of the
guaranteed time
  observations (GTO) program termed
  Mid-Infrared Characterization of Nearby Iconic galaxy Centers
(MICONIC) of the MIRI European
Consortium. The other targets in this
program are 
Mrk~231 \citep{AlonsoHerrero2024}, Arp~220 \citep[][and also van der 
Werf et al. in prep.]{Buiten2025}, NGC~6240
 \citep{HermosaMunoz2025},  
SBS0335-052, and the region surrounding
  SgrA$^*$. Two companion papers with MIRI-MRS observations of Cen~A study the
  distribution, temperature, and excitation of the warm molecular gas
  H$_2$ (Evangelista et al. in prep.) 
  and properties of the polycyclic aromatic hydrocarbons (PAH) features (Pantoni
  et al. in prep.).

The proximity of Cen~A  provides a unique
opportunity to study the impact of the radio jet on the ionized and
molecular gas, at physical scales  relatively
close to its launch region. Because Cen~A is crossed by a prominent dust
lane, most of the detailed morphological and kinematic studies
 of the gas in its central region (several hundreds of parsecs) have
utilized near-infrared (near-IR) and  (sub)millimeter
observations \citep[see e.g.,][]{Marconi2000, Neumayer2007,
  Krajnovic2007, Espada2009, Espada2017}.  Within the inner $\sim 20\arcsec$,
  the cold molecular 
gas is distributed in several components, including a circumnuclear
disk (CND) with a projected size of $20\arcsec \times 10\arcsec$, a
$9\arcsec \times 6\arcsec$ nuclear ring that  is approximately covered
by our MRS observations, nuclear 
filaments within this ring, and a nuclear disk of a few parsecs in size
\citep{Espada2017}. On these scales, the hot and cold molecular gas,
as traced with the H$_2$ J=1–0 S(1) $2.12\,\mu$m and CO(3-2)
transitions, respectively, shows
rotational motions together with complex
noncircular motions. The latter were attributed to different
processes, including the presence of a nuclear warped disk
\citep{Neumayer2007}  or streaming motions, which might be produced by a putative
nuclear bar \citep{Espada2017}. The emission of the near-IR high
excitation line [Si\,{\sc vi}] was found to be extended along the direction of the radio
jet, and its kinematics were explained by a backflow of gas accelerated by the
radio jet \citep{Neumayer2007}. Other mechanisms, such as a nuclear
outflow,  a warp, or non-axisymmetric perturbations,  were also
proposed to explain the kinematics of
the near-IR [Fe\,{\sc ii}] line \citep{Krajnovic2007}. Previous
Spitzer/IRS spectral mapping with limited angular resolution revealed
that the emission of mid-IR high ionization 
lines, such as, [O\,{\sc iv}] and [Ne\,{\sc v}], is extended over 20–25\arcsec \,
on each side of the AGN and at an orientation near the radio jet \citep{Quillen2008}
but had limited spatial resolution. With the high angular resolution and
sensitivity achieved with 
 MIRI-MRS, in this work we study Cen~A's gas kinematics in the mid-IR with physical
 resolutions of approximately 5-10\,pc.

  This paper is organized as follows. Section~\ref{sec:observations}
  describes
  the MIRI-MRS observations and data analysis. In
  Sect.~\ref{sec:results} we present the results for the nuclear
  emission, the extended emission, and the modeling of the molecular
  and ionized gas kinematics. We discuss and summarize the main results  in
  Sect.~\ref{sec:discussion}. 

  \begin{figure}

    \hspace{0.25cm}
 \includegraphics[width=8.5cm]{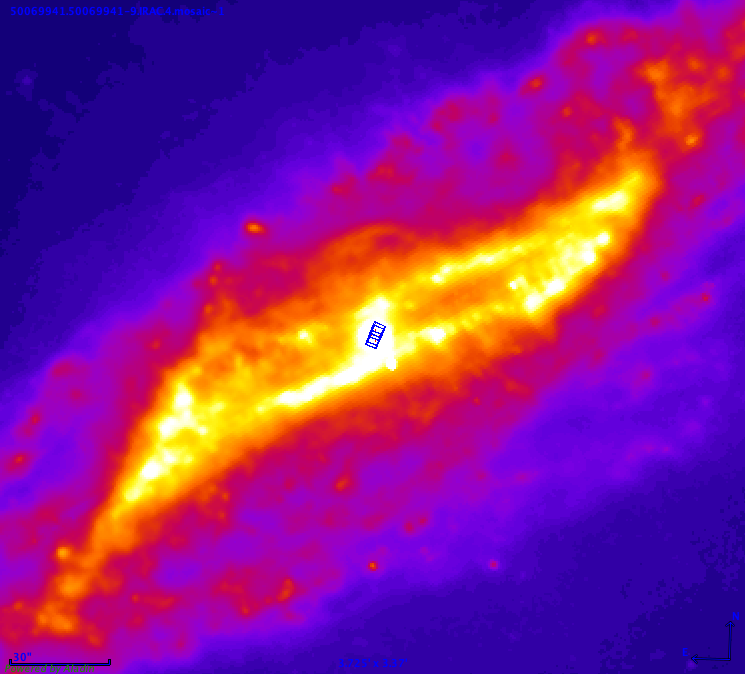}

    \caption{{\it Spitzer}/IRAC image at $8\,\mu$m of the 
      central $\sim 3.7\arcmin \times 3.4\arcmin$ region of Cen~A.
      The footprints of the MIRI-MRS observations (blue rectangles) are shown. 
      Orientation is north up, east to the left. Figure generated with
      the Astronomer's Proposal Tool (APT)
      version 2025.1.}
    \label{fig:SpitzerIRACimage}
  \end{figure}
  
\begin{figure*}
     \includegraphics[width=8.5cm]{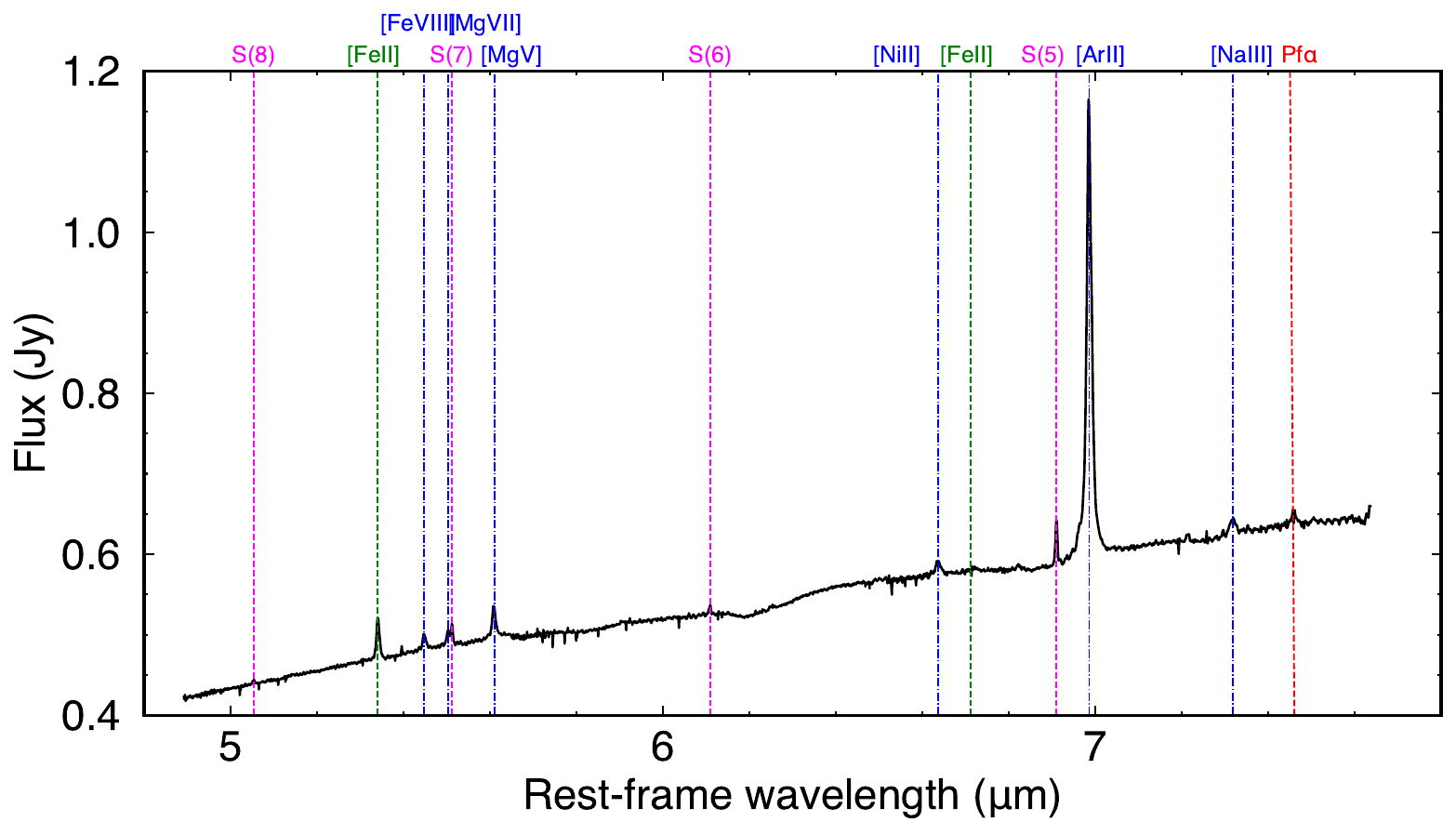}
\hspace{0.5cm}
     \includegraphics[width=8.5cm]{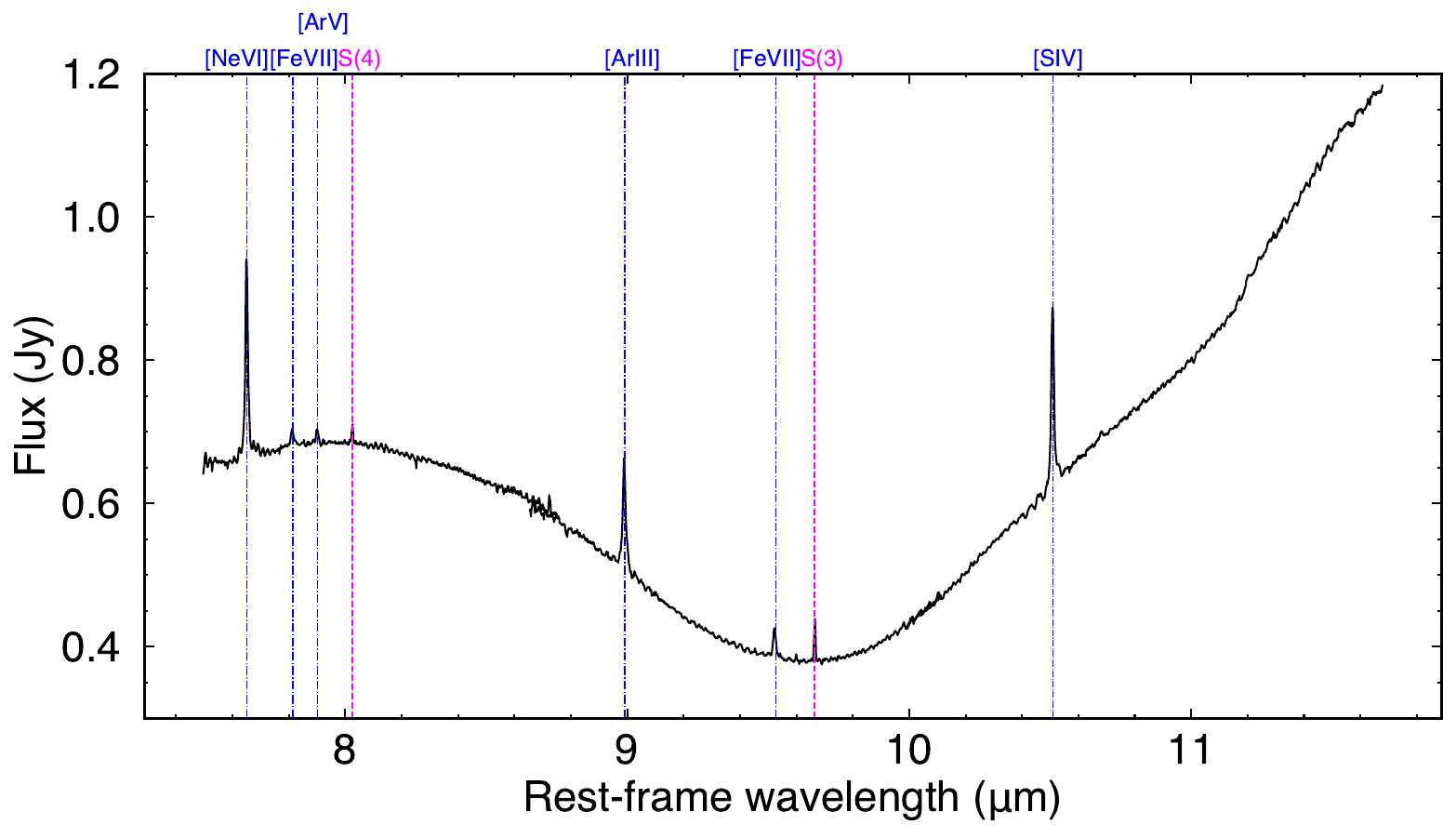}
     
     \includegraphics[width=8.5cm]{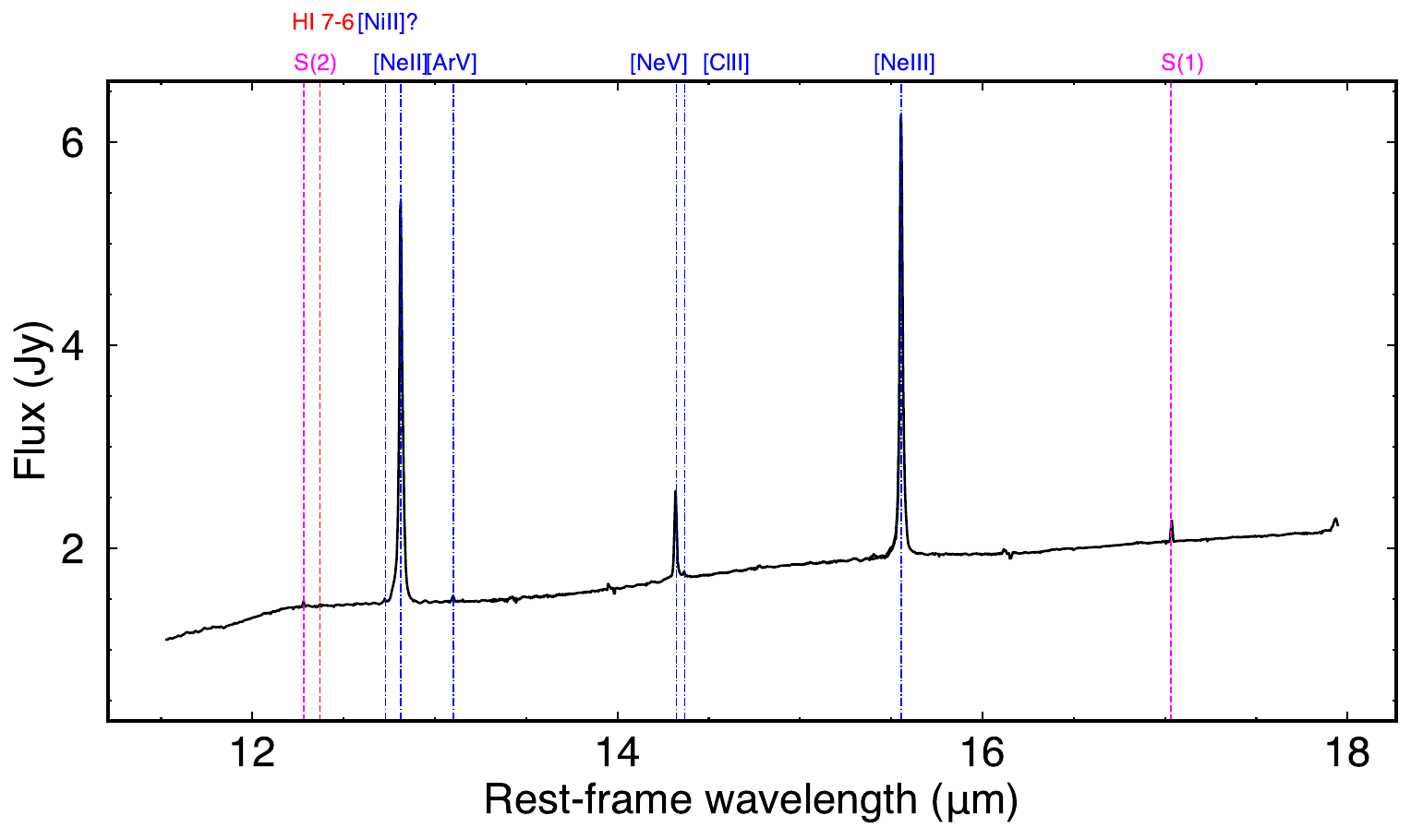}
\hspace{0.5cm}
     \includegraphics[width=8.5cm]{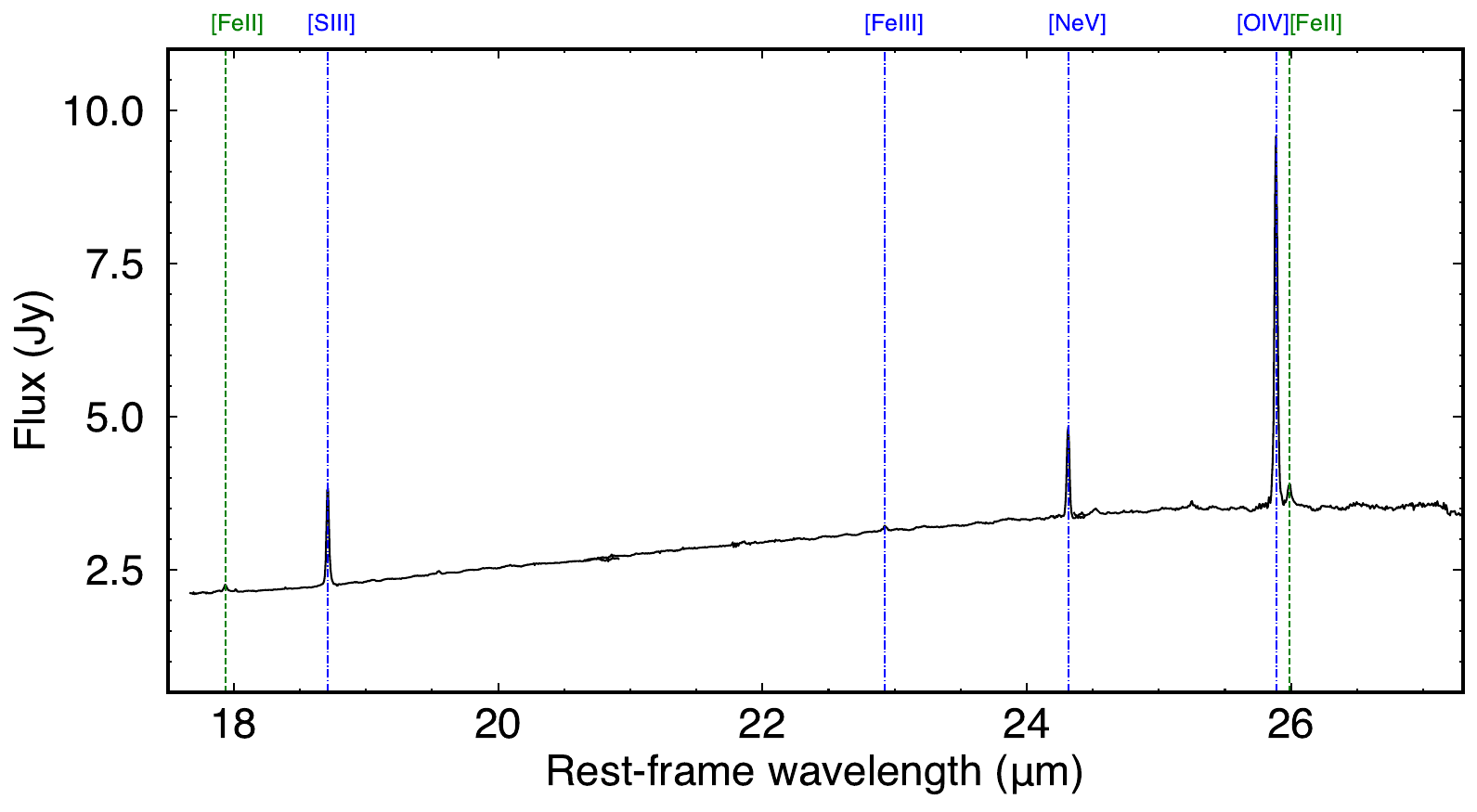}
     \caption{MIRI-MRS spectra of Cen A extracted as a point source
       for ch1 (upper left), ch2 (upper right), 
       ch3 (lower left), and ch4 (lower right). The lines mark fine-structure
       emission lines (blue and green), hydrogen recombination lines
       (red), and rotational H$_2$ 0--0 lines
       (magenta).}
     \label{fig:nuclearspectrum_channels}
\end{figure*}

\begin{table}
  \caption{Fine-structure lines.}
  \begin{tabular}{lccccc}

    \hline
    Line & $\lambda_{\rm rest}$ & IP & flux & 
                                                              FWHM$_{\rm
                                              line}$ & $\epsilon$\\
    \hline

    ${\rm [Fe\,II]}$ & 5.340 &  7.9 & $4.84\pm 0.11$ & 445 & 0.9\\ 
    ${\rm [Ar\,II]}$ &  6.985 & 15.8 & $49.55\pm 0.97$ & 608 & 5.7\\ 
    Pf$\alpha$ & 7.460 & ... & $1.08 \pm 0.11$ & 552 & 1.3\\
    ${\rm [Ne\,VI]}$ & 7.652 & 126.2 & $16.13 \pm 0.39$ & 397 & 3.8 \\
    ${\rm [S\,IV]}$  & 10.511 & 34.8 & $8.25 \pm 0.20$ & 335 & 1.3\\ 
    ${\rm [Ne\,II]}$ & 12.814 & 21.6 & $156.45 \pm 4.89$ & 539 & 6.5\\
    ${\rm [Ne\,V]}$ & 14.322 & 97.1& $19.80 \pm 0.47$ & 316 & 2.0\\
    ${\rm[Ne\,III]}$ &15.555 & 41.0 & $110.78 \pm 2.80$ & 396 & 37.5\\
    ${\rm [S\,III]}$& 18.713 & 23.3 & $33.39 \pm 0.88$ & 367 & 3.5\\
    ${\rm [Ne\,V}]$& 24.318 & 97.1 & $18.72 \pm 0.28$ & 286 & 1.4\\
    ${\rm [O\,IV]}$& 25.890 & 54.9 & $73.88 \pm 1.33$ & 307 & 1.8\\
    \hline
  \end{tabular}
  Notes.--- These are measured from spectra extracted as a point
    source (Sect.~\ref{subsec:analysis}). Wavelengths are in $\mu$m,
    IP in eV, fluxes in units of $10^{-14}$ erg cm$^{-2}$ s$^{-1}$
    and 
  line widths in km\,s$^{-1}$ are from fits with a single Gaussian.  
  FWHM$_{\rm line}$ are corrected for instrumental resolution. We only
  list lines analyzed in this work.
  \label{tab:nuclearfluxes}
  \end{table}

 \section{MIRI-MRS observations}\label{sec:observations}
\subsection{Data reduction}\label{subsec:datareduction}
The MIRI-MRS observations were taken as part of program ID 1269 covering
the full $\simeq 5-28\,\mu$m spectral range and
consisted of a $2\times 1$ mosaic (see Fig.~\ref{fig:SpitzerIRACimage}). We used a four
point dither pattern for extended sources in each of the mosaic
pointings, with a total integration time of 600\,s per mosaic pointing,
in FASTR1 readout mode. We also obtained background observations with
a two point dither, extended source pattern, with a 300\,s of total integration
time.  We reduced the MRS data using version 1.13.4 of
the JWST Science Calibration Pipeline \citep{Bushouse2023},
and the files from context 1185 of the Calibration References
Data System (CRDS). We followed the standard MRS pipeline procedure
\citep{Dicken2022,
  Morrison2023, Argyriou2023, Gasman2023, Patapis2024} for the galaxy
and background observations and produced
drizzled data cubes \citep{Law2023} with the default spaxel scales of
0.13, 0.17,  0.20, and 0.35\arcsec \, for channels 1, 2, 3, and 4 (ch1, ch2, ch3,
and ch4, hereafter).

During the pipeline reduction, we
switched off the background corrections and sky matching.
Following \cite{PereiraSantaella2022}, we created a background image per spectral element by
computing the median of all the spaxels of the fully reduced
background cubes, which we then subtracted from
the on-source cubes on a spaxel-by-spaxel basis. We generated the
reconstructed data cubes with  
the usual orientation of north up, and east to the left. The 
resulting fields of view (FoV) cover approximately the central
$\simeq  6.6\arcsec \times 8.5\arcsec$ to $13\arcsec \times
14.4\arcsec$ for ch1 and ch4,
respectively. In Fig.~\ref{fig:SpitzerIRACimage} we overlay the
ch4 FoV on the {\it Spitzer}/IRAC $8\,\mu$m
image showing the dusty parallelogram-shape structure \citep[see][]{Quillen2006}.

\subsection{Data  analysis}\label{subsec:analysis}
Cen~A shows a strong point-like continuum source
at the AGN position with a full width
at-half-maximum (FWHM) in ch1-short of 0.3\arcsec. Thus, to isolate
the nuclear unresolved emission, 
we extracted spectra using an aperture of  $2\times
{\rm FWHM}$ and applied the corresponding point source correction. The
resulting spectra are plotted separately for each channel in 
Fig.~\ref{fig:nuclearspectrum_channels} and the full spectrum in Fig.~\ref{fig:fullspectrum}. We only 
needed to apply a small 
scaling factor to stitch together the ch1-short portion with that of
ch1-medium. We marked the numerous fine
structure lines and 
  H$_2$ 0--0 S(1) to S(8) transitions 
  detected in the nuclear region of Cen~A. For the analysis of the excitation properties
of the ionized gas in the nuclear region of Cen~A
(Sect.~\ref{subsec:nuclearlineratios}) we fit the lines using only a single
Gaussian and a local continuum. The fluxes and FWHM of the fine (FWHM$_{\rm line}$)
structure lines analyzed in this work are presented in Table~\ref{tab:nuclearfluxes}.

\begin{figure*}
 \sidecaption
\includegraphics[width=12cm]{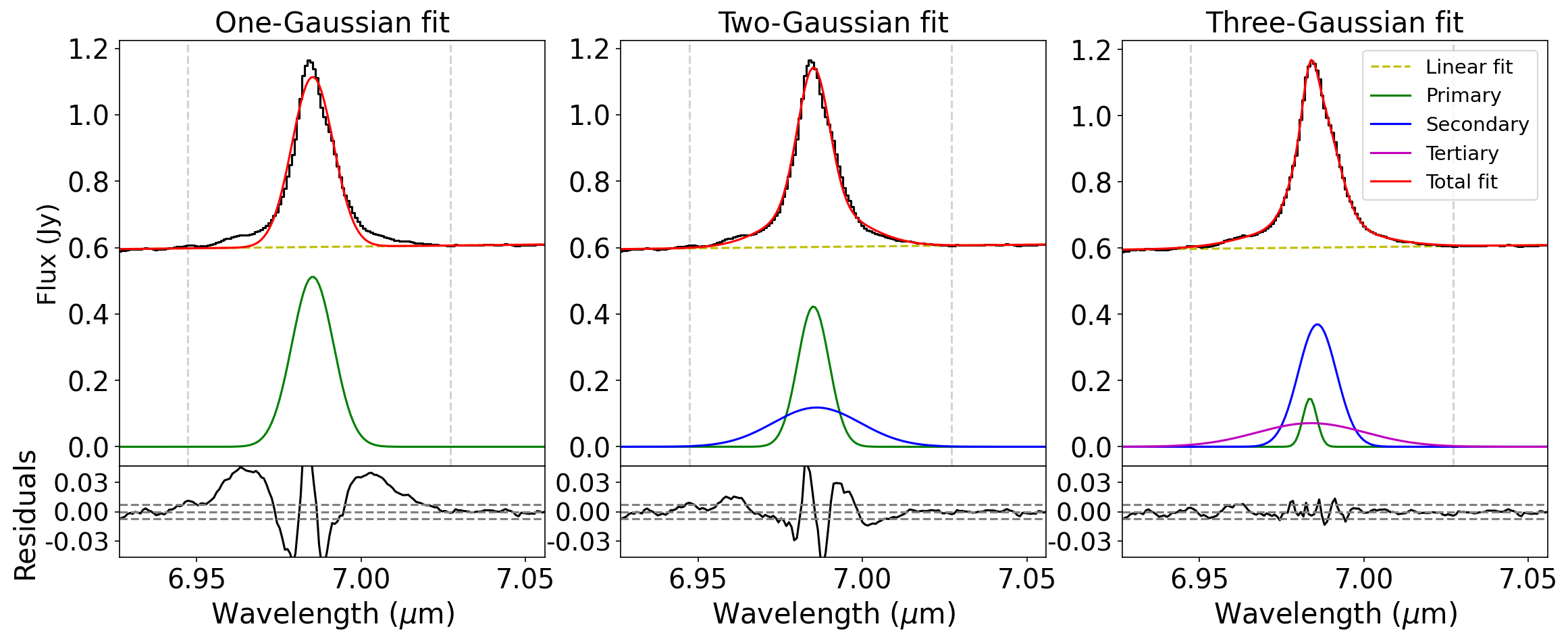}
\caption{Parametric fits of the [Ar\,{\sc ii}] line profile from
    the spectrum extracted as a point source. 
 Fits are (left to right) with
       one, two, and three Gaussians plus a local
       continuum. The top panel shows the total fit and  the individual
       Gaussians and the bottom the residuals of the fit. The vertical
       lines in the main panel indicate  the continuum regions and the
       horizontal lines in the bottom panel $\pm 3 \times s_{\rm
         cont}$, where $s_{\rm cont}$ is the local continuum
       standard deviation. The
       corresponding
       parameters of each fit are listed 
     in Table~\ref{tab:parametricfits}.}
     \label{fig:lineprofiles_ArII}
\end{figure*}

\begin{figure}
  \hspace{0.5cm}
  \includegraphics[width=8cm]{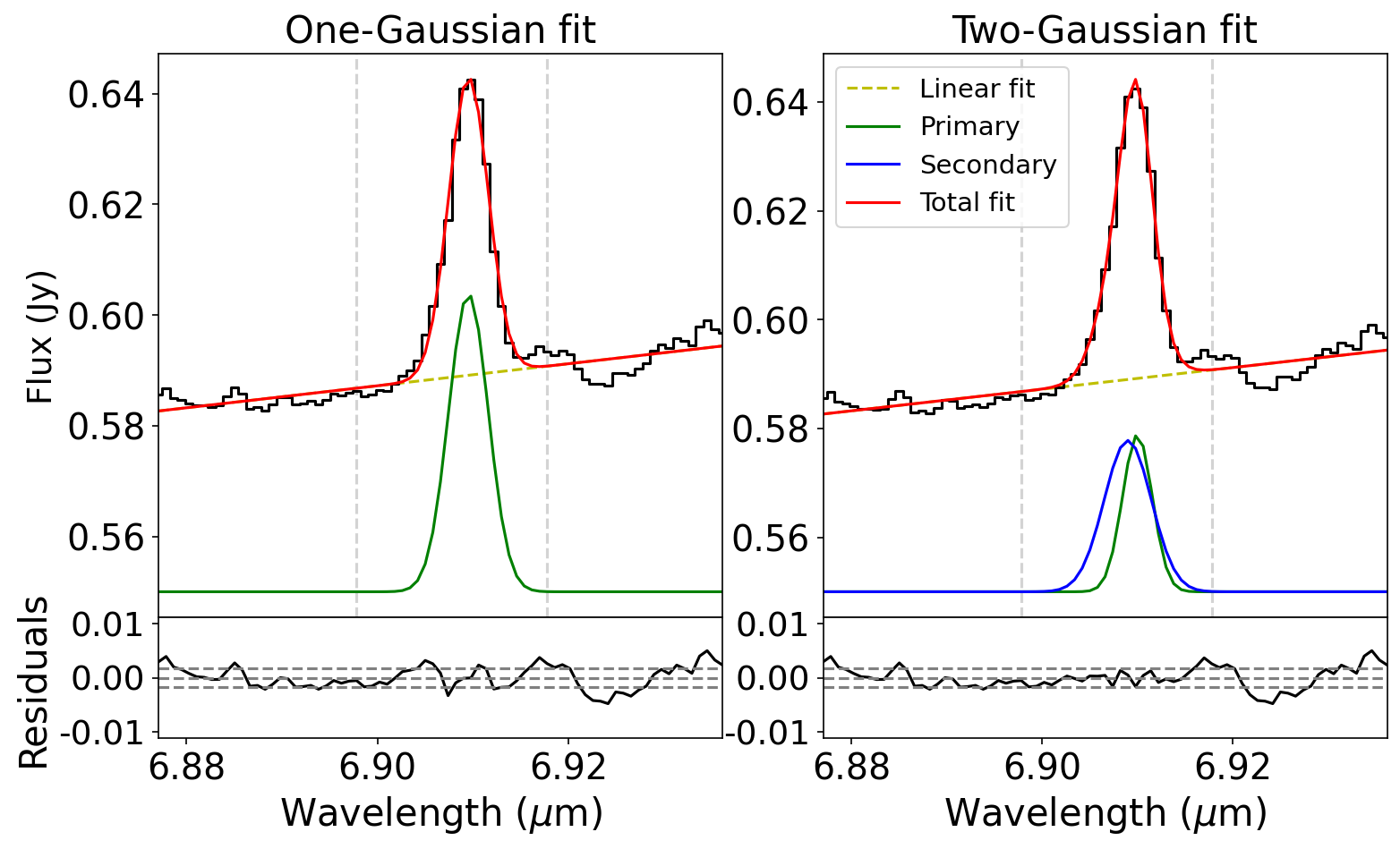}
     \caption{Same as Fig.~\ref{fig:lineprofiles_ArII}, but for
       H$_2$ S(5) and  fits only with one and two Gaussians (see text and
       Table~\ref{tab:parametricfits} for more details).}
     \label{fig:lineprofiles_H2S5}
\end{figure}

The fine-structure lines from the unresolved region are not fully
accounted for with a single Gaussian
component. We thus  performed additional fits using a local continuum and
a maximum of three Gaussian components. Details of the method
are discussed
   by \cite{HermosaMunoz2024LINERS, 
    HermosaMunoz2024NGC7172}. To 
  determine statistically the number of Gaussians needed to improve a fit, 
we computed the ratio between the standard deviations of the continuum underneath
the line after subtracting the
Gaussian modeling and that of the continuum adjacent to the line,
$\epsilon$. As reference, values of $\epsilon 
>3$ are generally taken as an indication that  
an extra component is needed for the fit
\citep[see][]{Cazzoli2018, 
  HermosaMunoz2024LINERS}. 
We show in Figs.~\ref{fig:lineprofiles_ArII} and
  \ref{fig:lineprofiles_H2S5} the results for the
  brightest lines in ch1 ([Ar\,{\sc ii}]
  and H$_2$ S(5) at $6.909\,\mu$)
  which has the highest spectral resolution as well as for ch3 line [Ne\,{\sc iii}]
  in  Fig.\ref{fig:lineprofiles_extra}.

\begin{figure*}

  \vspace{-0.4cm}
 \hspace{1.5cm} 
  \includegraphics[width=14.7cm]{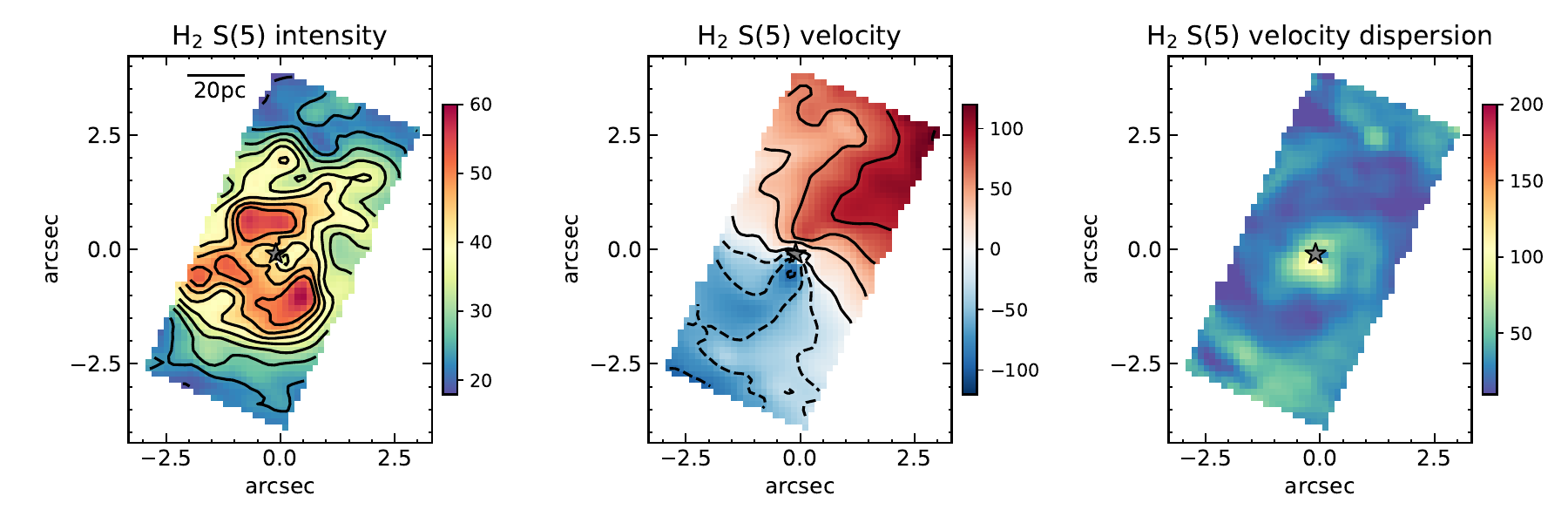}

    \vspace{-0.25cm}
 \hspace{1.5cm} 
  \includegraphics[width=14.7cm]{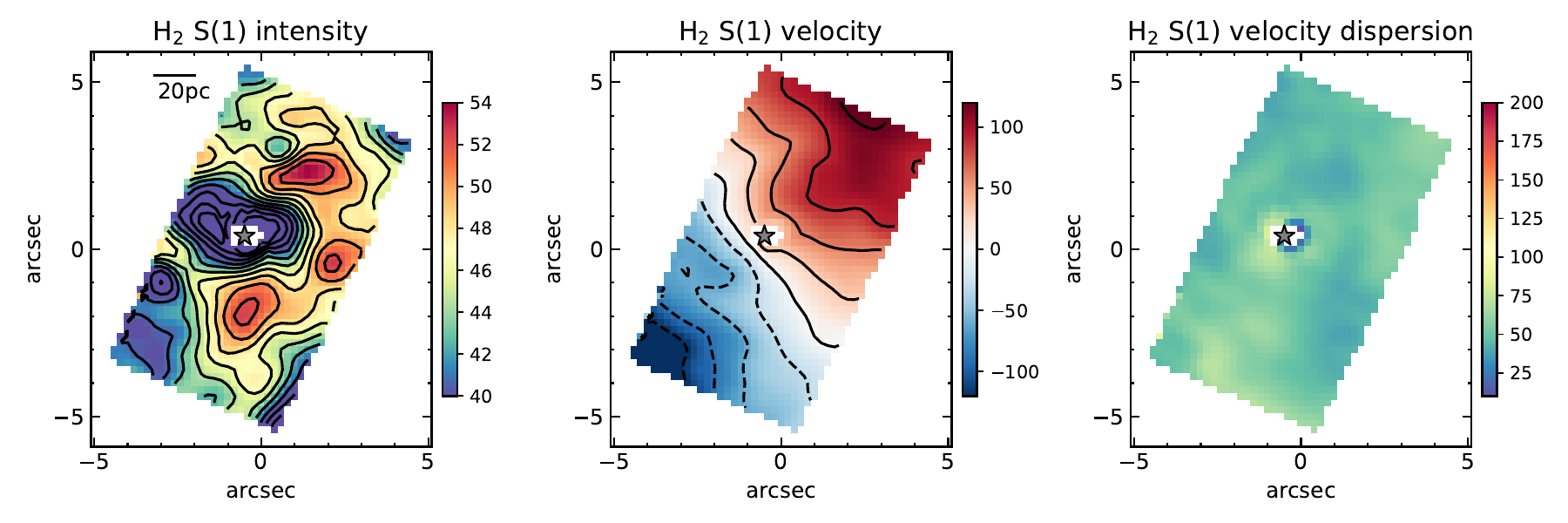}
     \caption{Maps of  H$_2$ S(5) at $6.909\,\mu$m (top) and
         H$_2$ S(1) at $17.03\,\mu$m (bottom)  constructed as
       explained in Sect.~\ref{subsec:analysis}. 
       Panels show  the intensity and contours in a square root
       scale in arbitrary units (left),
       the mean-velocity field in units of km\,s$^{-1}$ (middle), 
       and the velocity dispersion map $\sigma$ (corrected for instrumental
       resolution)
       in units of km\,s$^{-1}$
       (right). The velocity contours are in a linear
       scale (solid lines positive values and dashed lines negative
       values). The star symbol marks the peak of the continuum 
       adjacent to the line, that is, the AGN position. The
       0,0 point on the axes refers to the center of the corresponding sub-channel
       array. North is up and east to the left.}
       \label{fig:maintextalucinemapsH2}

\end{figure*}

\begin{figure*}
  \vspace{-0.25cm}
 \hspace{1.5cm} 
 \includegraphics[width=14.7cm]{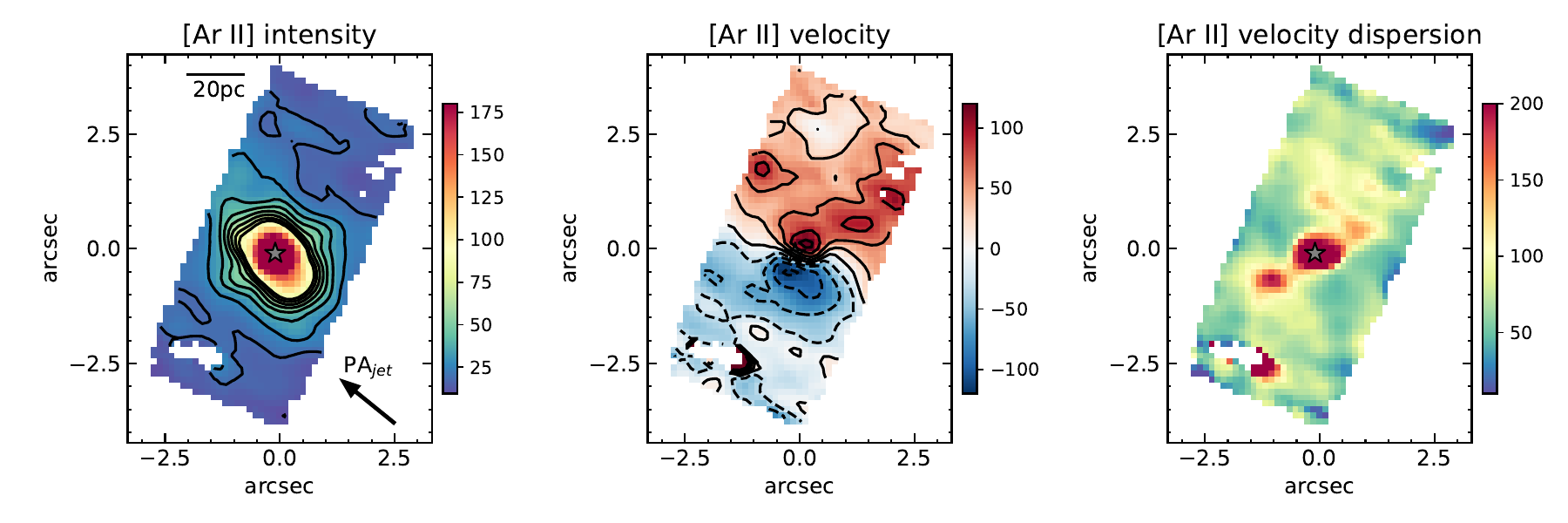}

  \vspace{-0.25cm}

 \hspace{1.5cm} 
 \includegraphics[width=14.7cm]{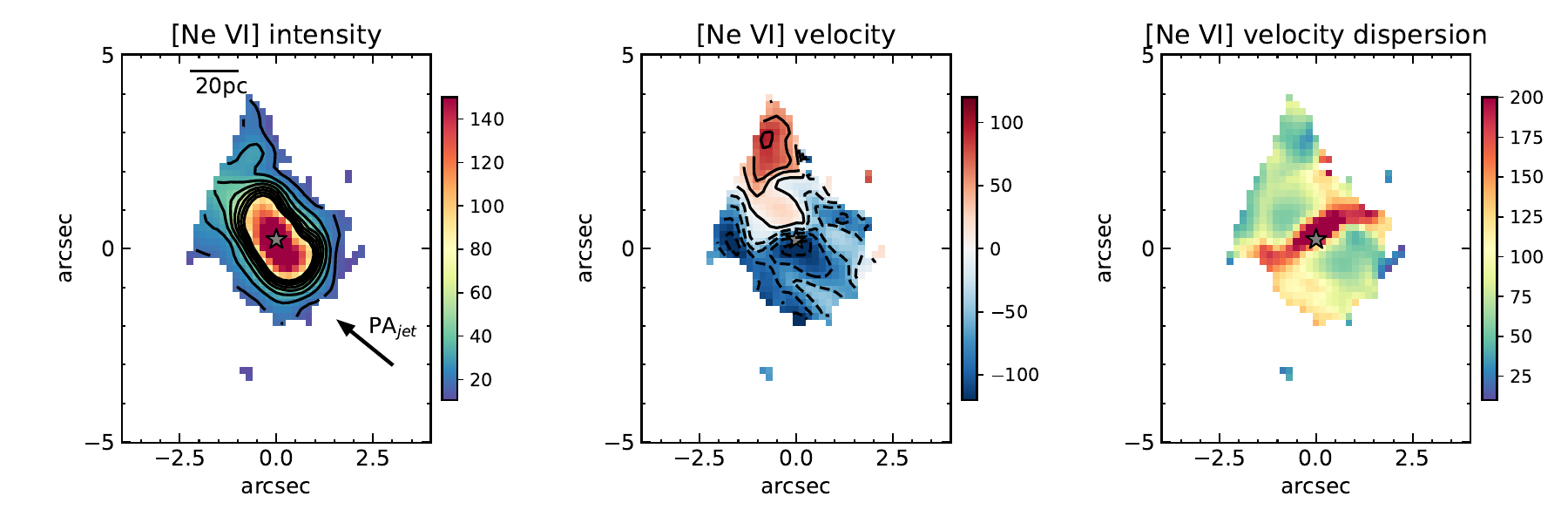}

  \vspace{-0.25cm}
 \hspace{1.5cm} 
  \includegraphics[width=14.7cm]{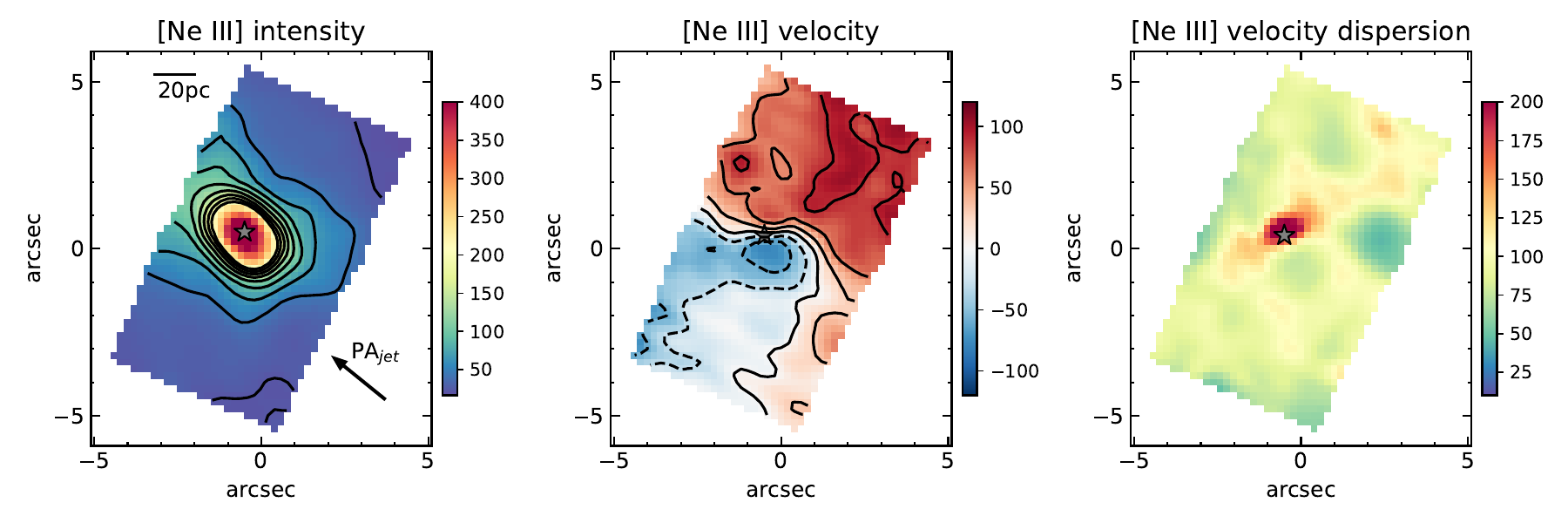}

  \vspace{-0.25cm}
 \hspace{1.5cm} 
  \includegraphics[width=14.7cm]{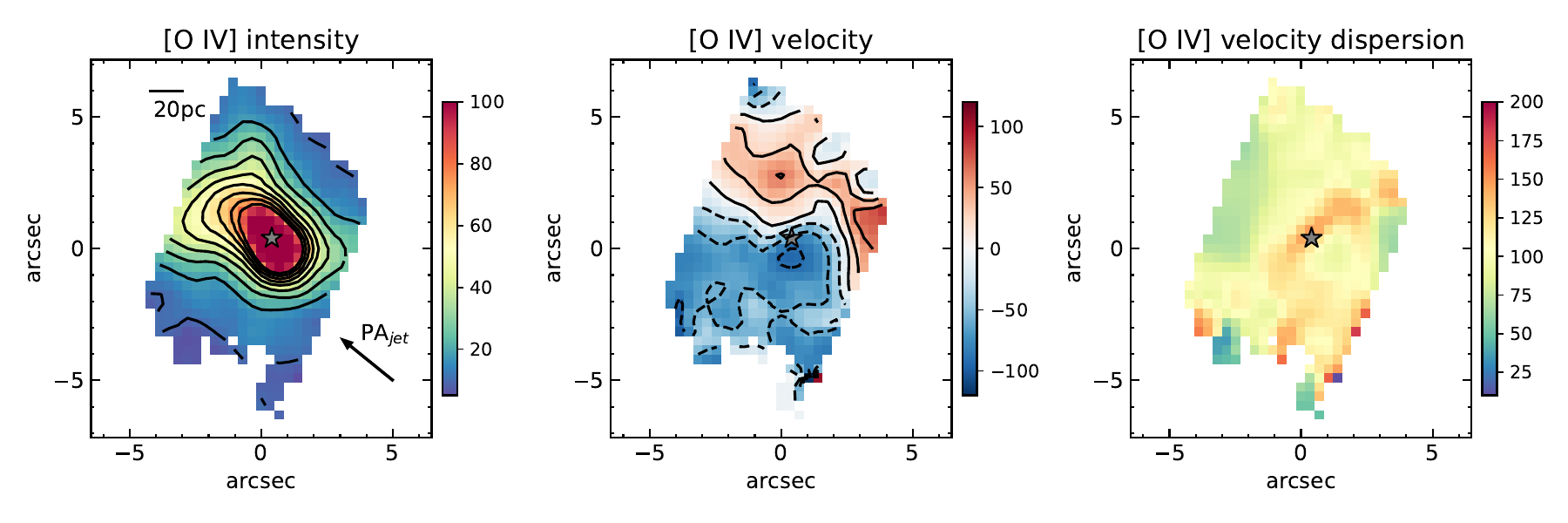}

  \vspace{-0.3cm}
\caption{Same as Fig.~\ref{fig:maintextalucinemapsH2} but for
    [Ar\,{\sc ii}], [Ne\,{\sc vi}],
       [Ne\,{\sc iii}], and [O\,{\sc iv}] (from top to
       bottom).}
     \label{fig:maintextalucinemapsFSL}
\end{figure*}

For the brightest fine structure lines and the
H$_2$ transitions we generated maps using the Python tool 
{\sc Alucine}
\citep{PeraltadeArriba2023}. This tool fits, on a spaxel-by-spaxel basis,  the selected line
using a user-specified number of Gaussian
components and a local continuum. We started by 
smoothing the fully reduced data cubes with a $2\times2$ spaxel average
box to increase the signal-to-noise ratio before fitting the
lines. For this analysis we used one and two Gaussians for the
fits and specified a threshold value of amplitude over noise 
of 3 for the line detection.  The outputs are maps of the fit line intensity, velocity
peak of the line (mean velocity field), and velocity dispersion. We
corrected the last quantity by subtracting in quadrature the
instrumental resolving powers 
($\sigma_{\rm inst}$) at
the corresponding line wavelength for ch1 and ch2
\citep{Argyriou2023} and the updated values for ch3
and ch4 \citep{Pontoppidan2024, Banzatti2025}. The
velocity fields are computed using a 
redshift value of  $z=0.001825$ or a corresponding value of the
systemic velocity of $v_{\rm sys}=547\,{\rm
  km\,s}^{-1}$. Maps produced with a single Gaussian for 
H$_2$ S(5) and H$_2$ S(1)  are in  Fig.~\ref{fig:maintextalucinemapsH2}
and for [Ar\,{\sc ii}], [Ne\,{\sc vi}],
[Ne\,{\sc iii}], 
and       [O\,{\sc iv}]
in Fig.~\ref{fig:maintextalucinemapsFSL}, while other
lines of interest, namely, [Fe\,{\sc ii}] at $\lambda_{\rm rest} =
5.34\,\mu$m,  [Ne\,{\sc ii}],
[S\,{\sc iv}], [Ne\,{\sc v}]
at $\lambda_{\rm rest} = 14.32\,\mu$m, and [S\,{\sc iii}] are 
in Fig.~\ref{fig:appendixalucinemaps}. We show the [Ar\,{\sc 
  ii}] maps  fit with two Gaussians in
Fig.~\ref{fig:appendixalucinemapstwocomponents}.

\begin{figure}
  \includegraphics[width=9cm]{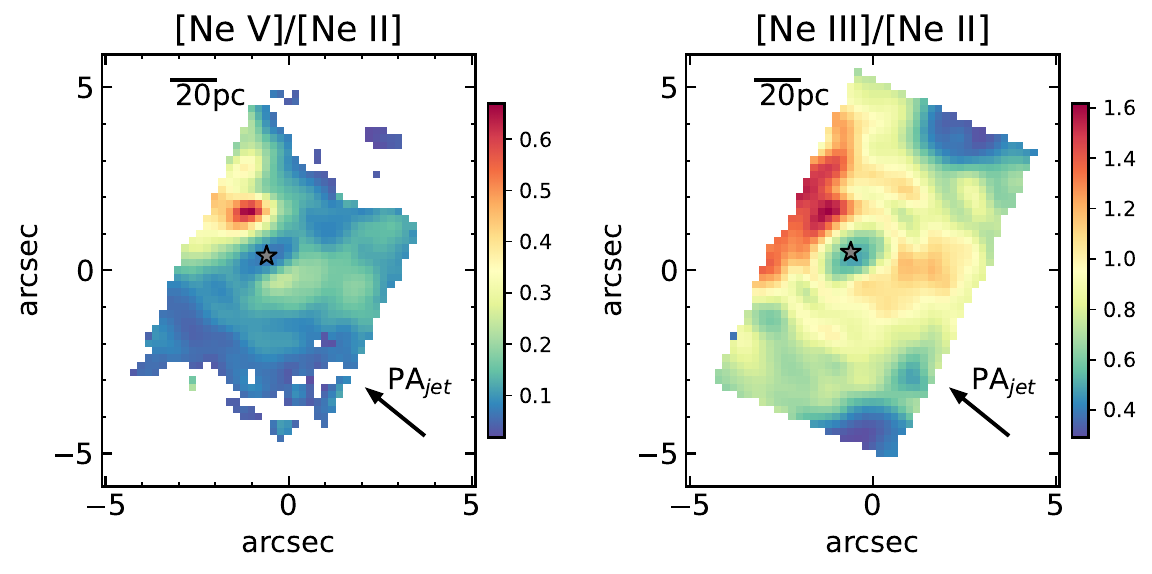}
  \caption{Maps of the [Ne\,{\sc v}]/[Ne\,{\sc ii}] (left)
       and [Ne\,{\sc iii}]/[Ne\,{\sc ii}] (right) ratios from the {\sc
         alucine} intensity maps computed with a single Gaussian. Some
    structure in the form of diffraction rings from the unresolved
    emission can be seen around the AGN position.  }
  \label{fig:lineratiomaps}
  \end{figure}

\section{Results}\label{sec:results}
\subsection{Line profiles}\label{subsec:nuclearemission}

For the line profiles from the spectra extracted as a point
  source (this section) and the
kinematics modeling (Sect.~\ref{subsec:kinematicmodeling}) of the fine
structure lines, we focus on [Ar\,{\sc ii}] 
and [Ne\,{\sc iii}]  because other  bright lines, such as, [Ne\,{\sc
  ii}], [Ne\,{\sc v}], [O\,{\sc iv}], have nearby features, that is, the $12.7\,\mu$m PAH
feature (although it is not detected or is faint in the nuclear
region), [Cl\,{\sc ii}]  at $\lambda_{\rm rest} = 14.37\,\mu$m,
and [Fe\,{\sc ii}] at $\lambda_{\rm rest} = 25.99\,\mu$m, respectively.

Fits with three Gaussians are  preferred to those with one or two
components for both [Ar\,{\sc ii}] and [Ne\,{\sc iii}], in terms of a visual inspection
of the residuals (Figs.~\ref{fig:lineprofiles_ArII} and \ref{fig:lineprofiles_extra}) 
and the $\epsilon$ parameter (see
Table~\ref{tab:parametricfits}). 
 The broadest components (corrected for instrumental
broadening) have FWHM$_{\rm line} \simeq
1425-1650\,{\rm km\,s}^{-1}$. To our knowledge, the recent work by
\cite{Marconcini2025} and this paper report, for the first time,
ionized gas broad lines  for Cen~A, which are likely associated with an outflow
driven by the radio-jet and/or the AGN (see
Sect.~\ref{subsec:outflowenergetics}). These FWHM values are larger
than in some Seyfert
galaxies, which have typically
FWHM$_{\rm line} \ll $1000$\,{\rm
  km\,s}^{-1}$ for the low ionization potential (IP) lines \citep{PereiraSantaella2022, AlvarezMarquez2023, 
  Armus2023} observed with MIRI MRS. On the other hand, 
such broad components are present in two local LLAGNs \citep{Goold2024},
several regions
close to the radio lobes (scales of $\sim 1\,$kpc) and along the
jet  (scales of hundreds of parsecs) in the radio
galaxy and Seyfert IC~5063 \citep{Dasyra2024} as well as the
  $z\sim 4$ luminous radio galaxy TNJ1338-1942 \citep{Roy2024}.

In contrast to the ionized gas, the nuclear H$_2$ S(5) line profile of
Cen~A is well fit with a
single  Gaussian with $\sigma=83\,{\rm km\,s}^{-1}$ (left panel
of Fig.~\ref{fig:lineprofiles_H2S5}). While adding an extra component
 improves slightly the fit, it would not be 
 deemed necessary according to the $\epsilon$ criterium.
A similar behavior for the line 
widths in Cen~A was already
noted in the near-IR  by \cite{Neumayer2007}  and has been observed in
 radio-loud AGNs with fast nuclear ionized outflows 
\citep[e.g., UGC~8782, ][]{CostaSouza2024} and the LLAGN NGC~4395
(Nandi et al. in prep.).

\subsection{Spatially resolved gas morphology and kinematics}\label{subsec:extendedemission}

In this section we describe the morphology and kinematics of the
extended emission  of the ionized and molecular gas.

\subsubsection{Results from single Gaussian fits}
The intensity maps of the
H$_2$ transitions and
the fine structure lines are strikingly different, as can be
seen from Figs.~\ref{fig:maintextalucinemapsH2} and ~\ref{fig:maintextalucinemapsFSL} (also
Fig.~\ref{fig:appendixalucinemaps}). The warm 
molecular gas traced by H$_2$ S(1) appears to be distributed in a
ring-like structure with some bright knots and is
approximately coincident with the nuclear ring  identified in cold
molecular gas, which has a size of $9\arcsec \times 6\arcsec$ 
\citep{Espada2017}. However, the higher excitation S(5) transition
bright clumps are mostly within the ring  seen in S(1), which might be
associated with the 2\arcsec-diameter nuclear disk \citep{Espada2017}. We refer the
reader to Evangelista et al. (in prep.) for a detailed analysis of the H$_2$ emission.
On the other hand, the brightest emission of the
ionized gas lines, from the coronal line [Ne\,{\sc vi}]
to the low IP [Ar\,{\sc ii}] line  (Fig.~\ref{fig:maintextalucinemapsFSL}), has an
ionization-cone-like morphology. It is    
clearly oriented in the direction of the radio jet on these scales
\citep[PA$_{\rm jet}=51^{\rm o}$, ][]{Clarke1992}, as found for other
lines in the near-IR \citep{Marconi2000, Krajnovic2007,
  Neumayer2007}. The [Fe\,{\sc ii}] emission (Fig.\ref{fig:appendixalucinemaps}, top)
shows bright emission coming from an unresolved source centered at the
AGN position and extended emission along the cone and galaxy disk.

The [O\,{\sc iv}] line emission map fills the entire 
MRS FoV along the jet  direction with
a projected size along the cone axis of approximately 6\arcsec \, or
100\,pc (see Fig.~\ref{fig:maintextalucinemapsFSL}). Some of the extended
[O\,{\sc iv}] and [S\,{\sc iii}] morphology to the north of the AGN 
position resembles the Pa$\alpha$ and Pa$\beta$ knots detected at approximately
2 and
3\arcsec \, away from the nucleus
\citep[see ][]{Marconi2000, Krajnovic2007}. The region mapped
with MRS is only a portion of a much larger ionized gas structure
observed in  {\it Spitzer}/IRS spectral maps of [O\,{\sc
  iv}] and [Ne\,{\sc v}] at $\lambda_{\rm rest} = 24.32\,\mu$m, which extends over
at least 45\arcsec \, or 765\,pc \citep{Quillen2008}.
Low IP  bright lines (e.g., [Ar\,{\sc ii}], [Ne\,{\sc
  iii}], [S\,{\sc iii}]) present additional
emission  in  the galaxy disk region, covering an approximate projected size of up to 11\,\arcsec
(185\,pc) along the
major axis. Again, this is only the inner part of the much
larger scale star-forming warped disk \citep{Quillen2008}.

\begin{figure}
  \hspace{0.5cm}
     \includegraphics[width=8cm]{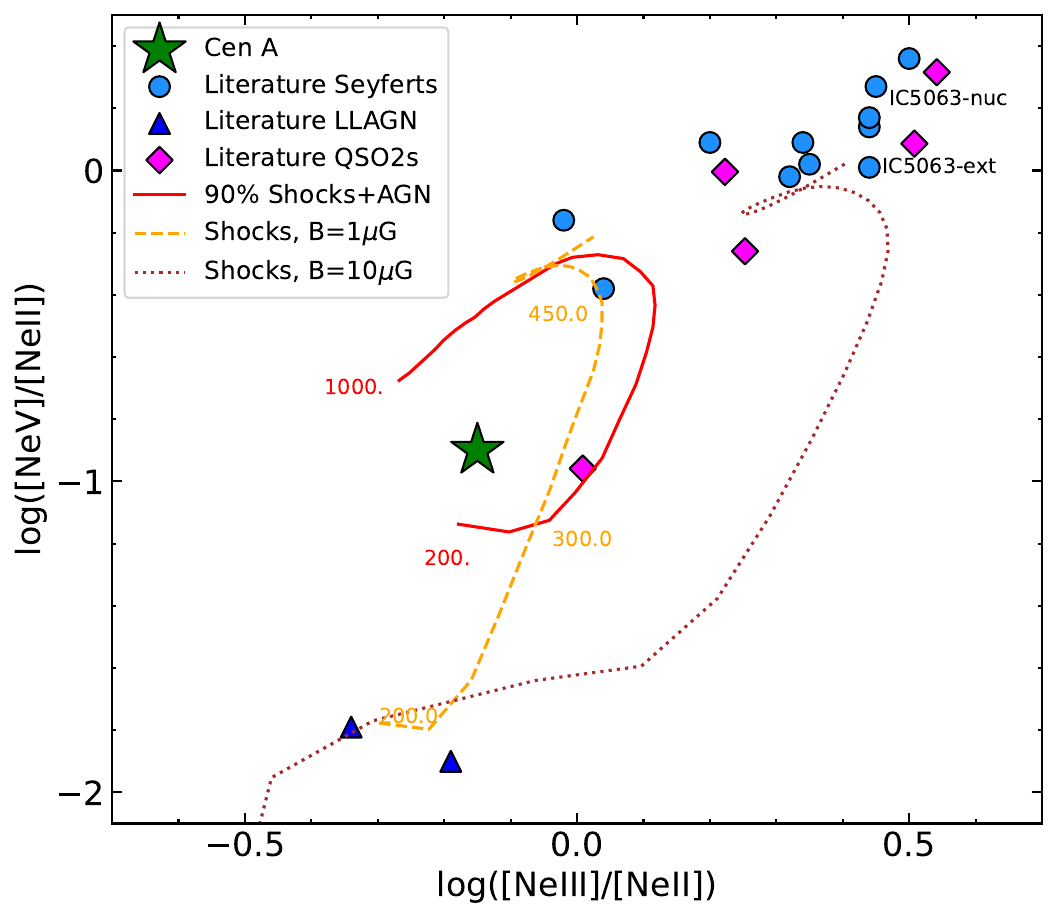}
     \caption{Diagnostic diagram [Ne\,{\sc v}]/[Ne\,{\sc ii}]
       versus  [Ne\,{\sc iii}]/[Ne\,{\sc ii}] comparing Cen~A's nuclear ratios
        with literature Seyferts, type 2 QSOs, and LLAGNs.
       For IC~5063  we  include the nuclear and integrated 2--3\,kpc ratios.
     The shock+AGN model track (solid red line) is from 
     \cite{Feltre2023} and the  two {\sc mappings v} shock models with precursor
     with the 
     parameters used in that work and two values of the transverse magnetic
       field of 1 and 10\,$\mu$G are from
     \cite{Alarie2019} (dashed orange and dotted brown
       lines, respectively). Numbers next to the curves indicate the
     shock velocities. See text for more details.}
     \label{fig:diagramratios}
\end{figure}

We also constructed [Ne\,{\sc v}]/[Ne\,{\sc ii}]
and [Ne\,{\sc iii}]/[Ne\,{\sc ii}] line ratio maps (see
Fig.~\ref{fig:lineratiomaps}).  At the AGN
position, the ratios are lower than in
the direction of the radio jet, including the Pa$\alpha$ and Pa$\beta$
knots. These  values are closer to those observed in some Seyfert
nuclei where AGN photoionization might dominate (see
Fig.~\ref{fig:diagramratios} and next section). This might indicate 
that several gas excitation mechanisms are at work. Along the inner part
of Cen~A disk along the galaxy 
major axis,  [Ne\,{\sc
  iii}]/[Ne\,{\sc ii}] varies between approximately 0.7 and 0.3, which are similar
to values observed in disks of nearby Seyfert galaxies
\citep{HermosaMunoz2024NGC7172, GarciaBernete2024}.

The mean-velocity fields of the ionized and warm molecular gas (middle panels of
Figs.~\ref{fig:maintextalucinemapsH2}, \ref{fig:maintextalucinemapsFSL}, and
\ref{fig:appendixalucinemaps}) display rotational motions, where the velocities
vary within a range of approximately  $-140\,{\rm
  km\,s}^{-1}$  to
$+120\,{\rm  km\,s}^{-1}$, depending on the line. There are however
strong deviations from rotation, which are modeled in Sect.~\ref{subsec:kinematicmodeling}.  The
velocity dispersion maps are markedly different for the ionized gas and
the warm H$_2$ gas, both in terms of the values and the
morphologies. The ionized gas velocity dispersion peaks at the AGN location and is
increased in the direction perpendicular to the radio jet
(Figs.~\ref{fig:maintextalucinemapsFSL} and 
\ref{fig:appendixalucinemaps}, right panels), as observed
in several Seyfert galaxies in the optical \citep[see e.g.,][]{Venturi2021} and
in the mid-IR with MRS \citep[see e.g.,][]{Davies2024, Zhang2024}. The
H$_2$ lines  (see Fig.~\ref{fig:maintextalucinemapsH2}) show
lower values of the velocity dispersion, with several spaxels in the
H$_2$ S(5) map having values close to the instrument spectral
resolution $\sigma_{\rm inst} \simeq 35\,{\rm km\,s}^{-1}$ for this
line\footnote{To produce the $\sigma$ maps of H$_2$ S(5) and 
for [Ar\,{\sc ii}], especially when using two Gaussians (next section)
for the fit, we artificially
  decreased slightly the instrumental resolution, $\sigma_{\rm inst}$, to avoid corrected $\sigma$ values (after
  subtracting  $\sigma_{\rm inst}$ in quadrature) of nearly
  zero in 
  those spaxels where the line width is close to the instrumental
  value.}. Additionally, the H$_2$ S(5) velocity dispersion map shows
a spiral-like morphology, which might be an indication of material
inflowing into the central region \citep[see 
e.g.][and Evangelista et al. 2025]{Maciejewski2004}.  On the other
hand, the regions with low 
velocity dispersion values are approximately coincident with the
bright H$_2$ S(5) knots.

\subsubsection{Results from two Gaussian fits}
As explained in Sect.~\ref{subsec:analysis}, we also produced
  spatially resolved maps fitting the lines with two Gaussians.
In general, interpreting this kind of fits can be challenging, since the
ordering and classification of the components are not always
straightforward. Nevertheless, for each spaxel we  ordered them
  in terms of the line width, with the first component corresponding to the narrow component and the second to the
broad one. Focussing on [Ar\,{\sc ii}]
(Fig.~\ref{fig:appendixalucinemapstwocomponents}), 
  the maps reveal that the central emission of the broad
component resembles the shape of this channel point source. This
suggests that this component, which has typical values of $\sigma
\simeq 230-330\,{\rm km\,s}^{-1}$,  originates mostly from an unresolved
region. At the MRS ch1-long
spatial resolution, this corresponds to a FWHM of approximately 0.35\arcsec \, or
6\,pc for the assumed distance to Cen~A. The velocity dispersion
  of the [Ar\,{\sc ii}] narrow component retains increased values in regions
  perpendicular to the radio jet. The results for
  [Ne\,{\sc iii}] (not shown) are similar, although the broad component
appears to be more extended than in [Ar\,{\sc ii}]. 

\begin{figure}
  \hspace{0.5cm}
     \includegraphics[width=8cm]{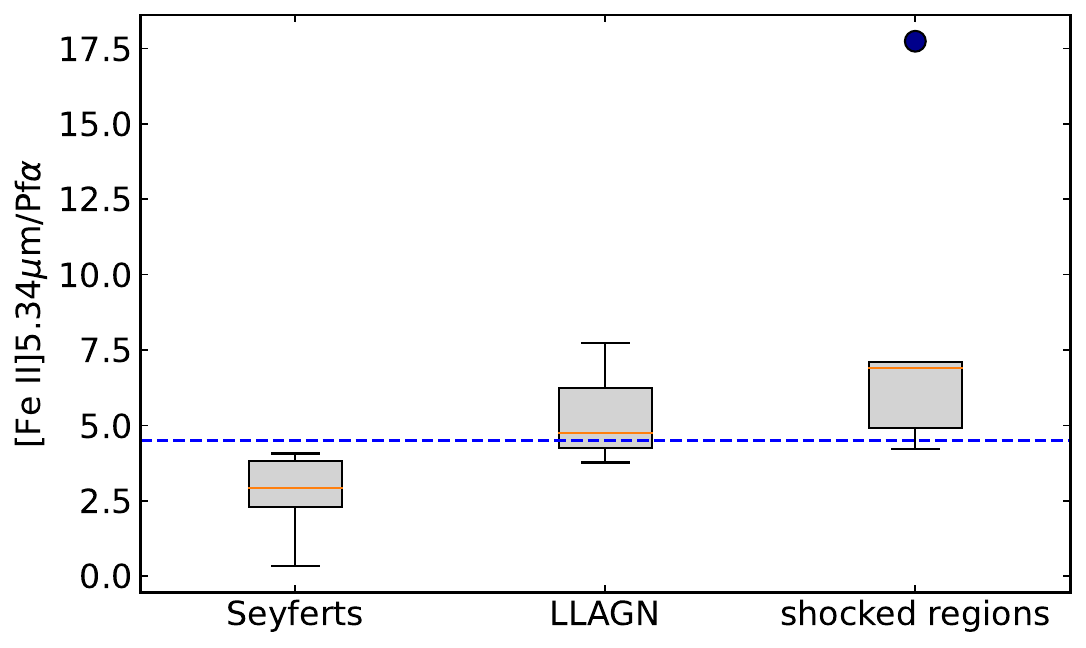}
     \caption{[Fe\,{\sc ii}]/Pf$\alpha$ box plot for Seyfert nuclei,
     LLAGNs including Cen~A's nuclear region, and shocked regions. The boxes comprise
     values from the
     first to the third quartiles of the distribution, the line inside the box is the
     median value, and the vertical lines contain 99.3\%  of the distribution.
     The outlier point
     in the shocked region class is the N2 jet impacted region of
     NGC~7319 \citep{PereiraSantaella2022}. The blue dashed line
       represents the nuclear value for Cen~A. The rest of references for the literature data are in text.}
     \label{fig:diagramratios_feiitopfalpha}
\end{figure}

\subsection{Mid-IR line ratios at the AGN position}\label{subsec:nuclearlineratios}
The nature of the excitation mechanism(s) of the ionized gas in the
central region 
of Cen~A has been investigated in several works. The NASA/IPAC
Extragalactic Database (NED) gives an optical 
classification as a Seyfert 2, while 
optical and near-IR line ratios of the nuclear region
seem to indicate low ionization nuclear emission line region (LINER) activity \citep{StorchiBergmann1997,
  Simpson1998}. On the other hand, \cite{Bicknell2013} showed
that the near-IR  ratios of high IP lines of a 
blueshifted  cloud located a few parsecs away from Cen~A nucleus
\citep[identified by][see also the discussion in
Sect.~\ref{subsec:kinematicmodeling}]{Neumayer2007} cannot be explained with shocks alone and require photoionization
by X-ray and UV emission arising from the AGN. We make
use of  several mid-IR line ratios to assess the ionization
mechanism in Cen~A's central region.

Figure~\ref{fig:diagramratios} shows a [Ne\,{\sc v}]/[Ne\,{\sc ii}]
versus  [Ne\,{\sc iii}]/[Ne\,{\sc ii}] diagram comparing Cen~A's
nuclear line ratios  (i.e., those measured in the spectra
  extracted as a point source, Table~\ref{tab:nuclearfluxes}) 
  together with
literature observations for Seyfert nuclei \citep{PereiraSantaella2022, AlvarezMarquez2023, Armus2023,
  HermosaMunoz2024NGC7172, Zhang2024}, radio-quiet type 2 QSOs
\citep{RamosAlmeida2025},  and  two LLAGNs
\citep[namely, NGC~1052 and the Sombrero galaxy, ][]{Goold2024}. We also included the line ratios   
 integrated over the inner $2-3\,$kpc of the radio galaxy
 IC~5063 \citep{Dasyra2024}, as
 well as our own
 measurements of the ratios of its nuclear region (log([Ne\,{\sc v}]/[Ne\,{\sc ii}])$=$0.27
 and log([Ne\,{\sc iii}]/[Ne\,{\sc ii}])$=$0.45). This figure
clearly demonstrates that Cen~A's  nuclear line
ratios are intermediate between Seyfert and type 2 QSO
nuclei\footnote{We note that the physical scales of the nuclear regions 
 of these are much larger (at least a factor
 of 10 in linear scale) than those probed for Cen~A.}  and those of LLAGNs.
For reference, the outflow regions of some Seyfert galaxies
present lower [Ne\,{\sc
  v}]/[Ne\,{\sc ii}] and [Ne\,{\sc iii}]/[Ne\,{\sc ii}] than the
nuclear regions \cite[see][]{PereiraSantaella2022, HermosaMunoz2024NGC7172, Zhang2024}. 
This may indicate an extra  contribution from shocks to the gas
excitation. Conversely, the increased neon line ratios in the cone region
of Cen~A compared with the nuclear values
(see Fig.~\ref{fig:lineratiomaps}) suggests that AGN
photoionization might have a higher contribution there.

Notably, the two radio galaxies IC~5063 and Cen~A do not present 
similar neon line ratios, which would suggest different
gas excitation mechanisms. 
\cite{Dasyra2024} concluded that overall the jet has a more important contribution than the
AGN radiation for the excitation
of the ionized gas in IC~5063, although they did not rule out
that the latter mechanism might have a role in the gas
excitation in regions previously cleared by the passage of the jet
and/or regions being directly impacted by the radio jet. 
However, Cen~A and IC~5063 have  $L_{\rm bol}$(AGN)/$P_{\rm
  jet}$ values of $0.5-4$ (see Sect.~\ref{sec:introduction}) and
$9-16$ \citep{Morganti2015}, respectively, which might explain the different neon
line ratios. We also note that Cen~A is variable in X-rays with
a  long-term mean AGN bolometric luminosity of
$10^{43}\,{\rm erg\,s}^{-1}$, although in the 2015-2020 period it has
been in a low luminosity
state \citep[see Fig.~5 of][]{Borkar2021}. 
 
We also display in Fig.~\ref{fig:diagramratios}
the AGN+shock model track 
from \cite{Feltre2023}. These models have a 90\% contribution to H$\beta$ 
from fast shocks  with velocities from 200 to $1000\,{\rm
    km\,s}^{-1}$. The shock models were computed by \cite{Alarie2019} 
  with the {\sc mappings}
  code \citep[version {\sc v},][]{Allen2008} for a  metallicity 
$Z=0.017$, a pre-shock density of $n_{\rm H} = 10^2\,{\rm cm}^{-3}$, and
a transverse magnetic field strength of $1\,\mu$G, and
include the contribution of a precursor H\,{\sc ii} region.  We also plot  shock models
alone \citep{Alarie2019} corresponding to this value of the magnetic
field as well as $10\,\mu$G. As can be seen from this figure, the 
shock track with the higher magnetic field extends to the region
occupied by Seyfert galaxies.  For reference, the extensive
AGN photoionization model grid computed
by \cite{Feltre2023} covers most of the region in
  this diagram with 
log([Ne\,{\sc iii}]/[Ne\,{\sc ii}])$>$0 and log([Ne\,{\sc
  v}]/[Ne\,{\sc ii}])$>$$-0.5$. Cen~A's nuclear line
ratios are closer to models with an
important shock contribution than local
Seyferts. One relevant unknown factor in this comparison between observations
and shock models with precursor 
is the strength of the transverse magnetic field. 

Enhanced near-IR [Fe\,{\sc ii}]  to hydrogen recombination
line ratios have been used to identify shocked regions in galaxies \cite[see
e.g.,][]{AlonsoHerrero1997, RodriguezArdila2005, Colina2015}, which
could be associated
with supernova remnants or radio jets.  We compiled MRS [Fe\,{\sc
  ii}]5.34$\mu$m/Pf$\alpha$ line ratios from the literature for Seyfert
nuclei \citep{PereiraSantaella2022, AlvarezMarquez2023, Zhang2024,
  Dasyra2024}, LLAGNs \citep{Goold2024}, and circumnuclear shocked
regions identified in NGC~7469
\citep{U2022} and NGC~7319
\citep{PereiraSantaella2022}. Figure~\ref{fig:diagramratios_feiitopfalpha}
presents the statistical results, where we included the nuclear ratio
of Cen~A ([Fe\,{\sc
  ii}]5.34$\mu$m/Pf$\alpha$=4.5, see Table~\ref{tab:nuclearfluxes})
in the LLAGN category. As can be seen from this figure, the nuclear
ratio of Cen~A appears to be closer to LLAGN and shocked regions than
Seyfert nuclei, as infered from the neon line ratios.

From this empirical comparison based on mid-IR line ratios, we conclude
that both shocks and AGN photoionization contribute to the gas excitation in the
nuclear region and along the jet direction of Cen~A in the inner $\sim$100\,pc, although the
former mechanism might have a more relevant
role when compared with some local
Seyfert nuclei.

\begin{figure}
\hspace{0.25cm}
  \includegraphics[width=8.5cm]{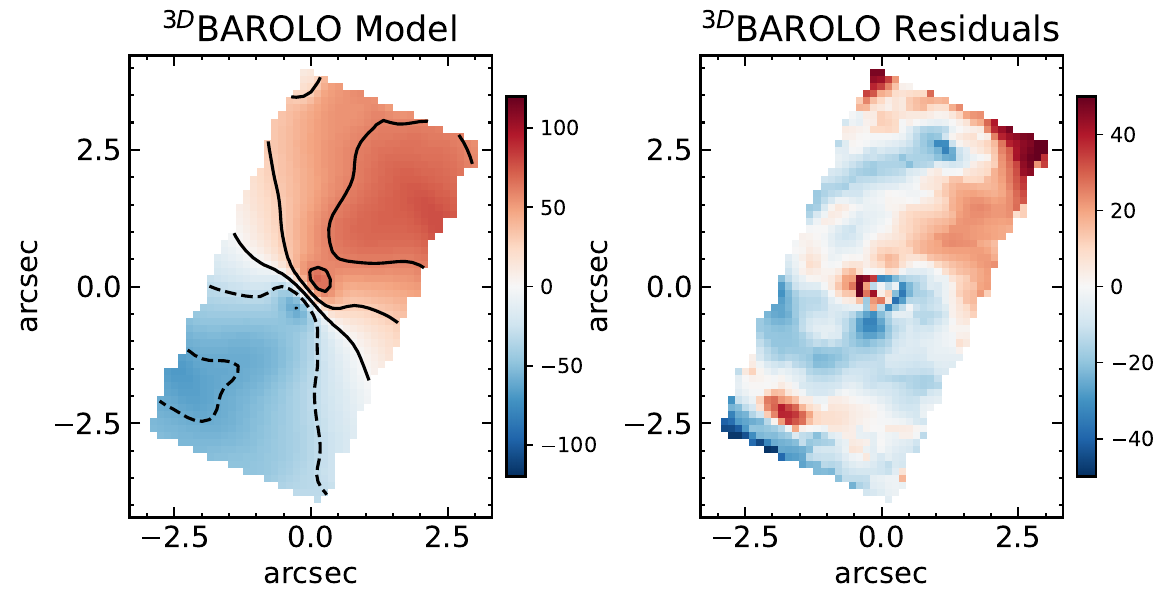}

\hspace{0.25cm}
  \includegraphics[width=8.5cm]{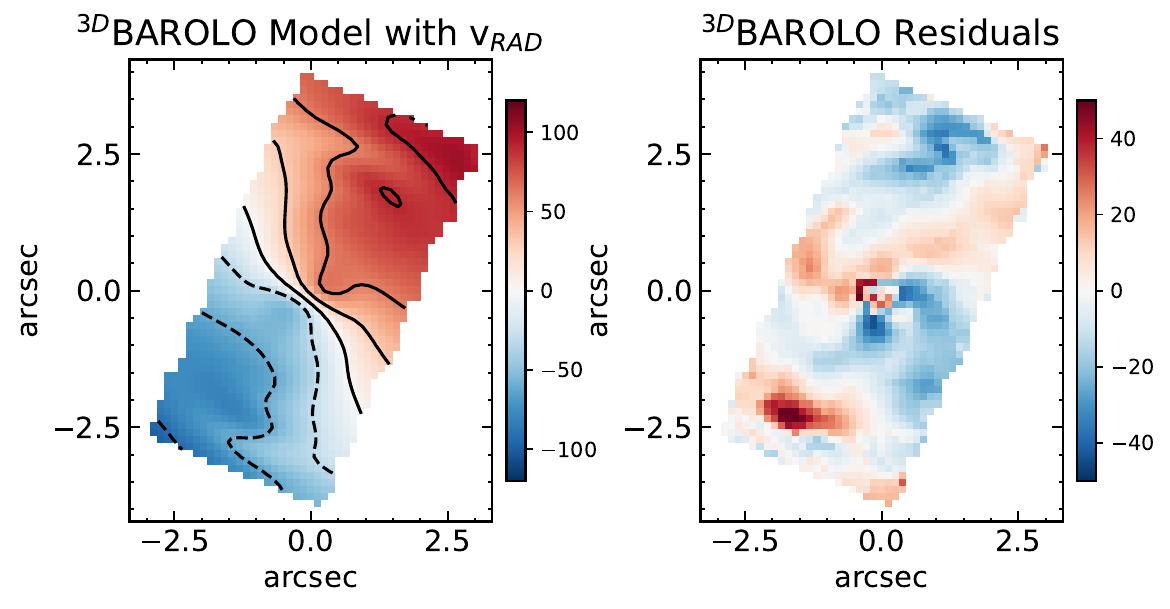}
     \caption{Top: Velocity field of the  $^{\rm 3D}$BAROLO rotating
       disk model (left) fit to the
       H$_2$ S(5) transition and residuals computed by subtracting
     the model from the $^{\rm 3D}$BAROLO moment 1 map (right). Bottom: Same as top 
     but including a $v_{\rm RAD}$ component in the
     model. Orientation, color scale, and contours for the velocity models are as in
     Fig.~\ref{fig:maintextalucinemapsH2}.} 
     \label{fig:BAROLOvfields_H2S5}
\end{figure}

\begin{figure}
  \hspace{1cm}
  \includegraphics[width=7cm]{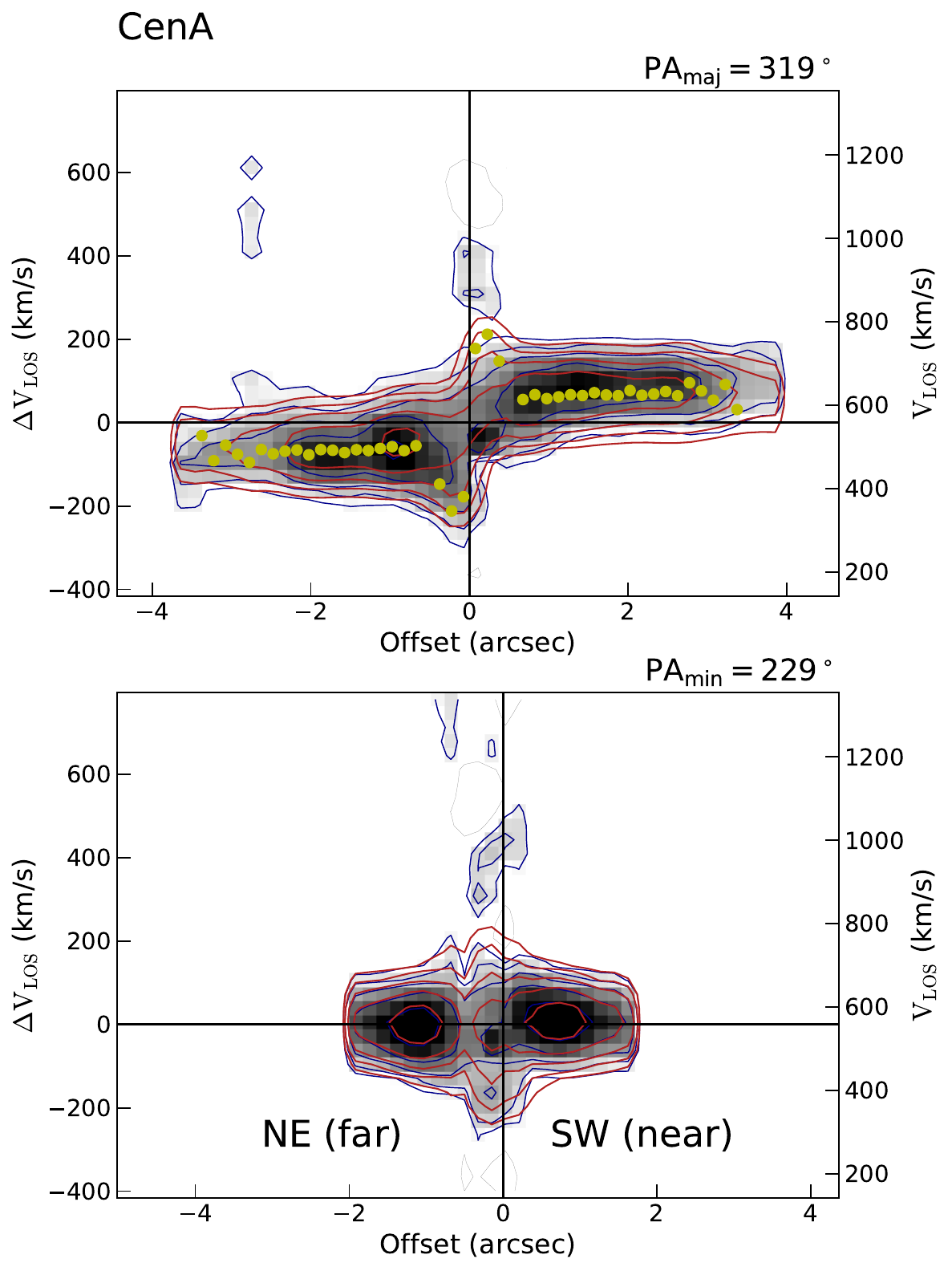}
\caption{H$_2$ S(5) p-v diagrams extracted along the kinematic major
  (top) and minor axes (bottom). The gray scale and blue 
  contours are the observations, and  red contours represent the fit $^{\rm
    3D}$BAROLO rotating disk model. Yellow dots show
  the fit rotation curve. In the central $\pm 0.5\arcsec$,
       some of the blueshifted velocity structure might be due to
       residuals left from subtracting the AGN strong
       continuum. PA$_{\rm maj}$ is measured  from north, 
  counterclockwise, and on the receding half of the galaxy disk. } 
     \label{fig:BAROLOpv_H2S5NOVRAD}
\end{figure}

\begin{figure}[hbt!]
  \hspace{1cm}
  \includegraphics[width=7cm]{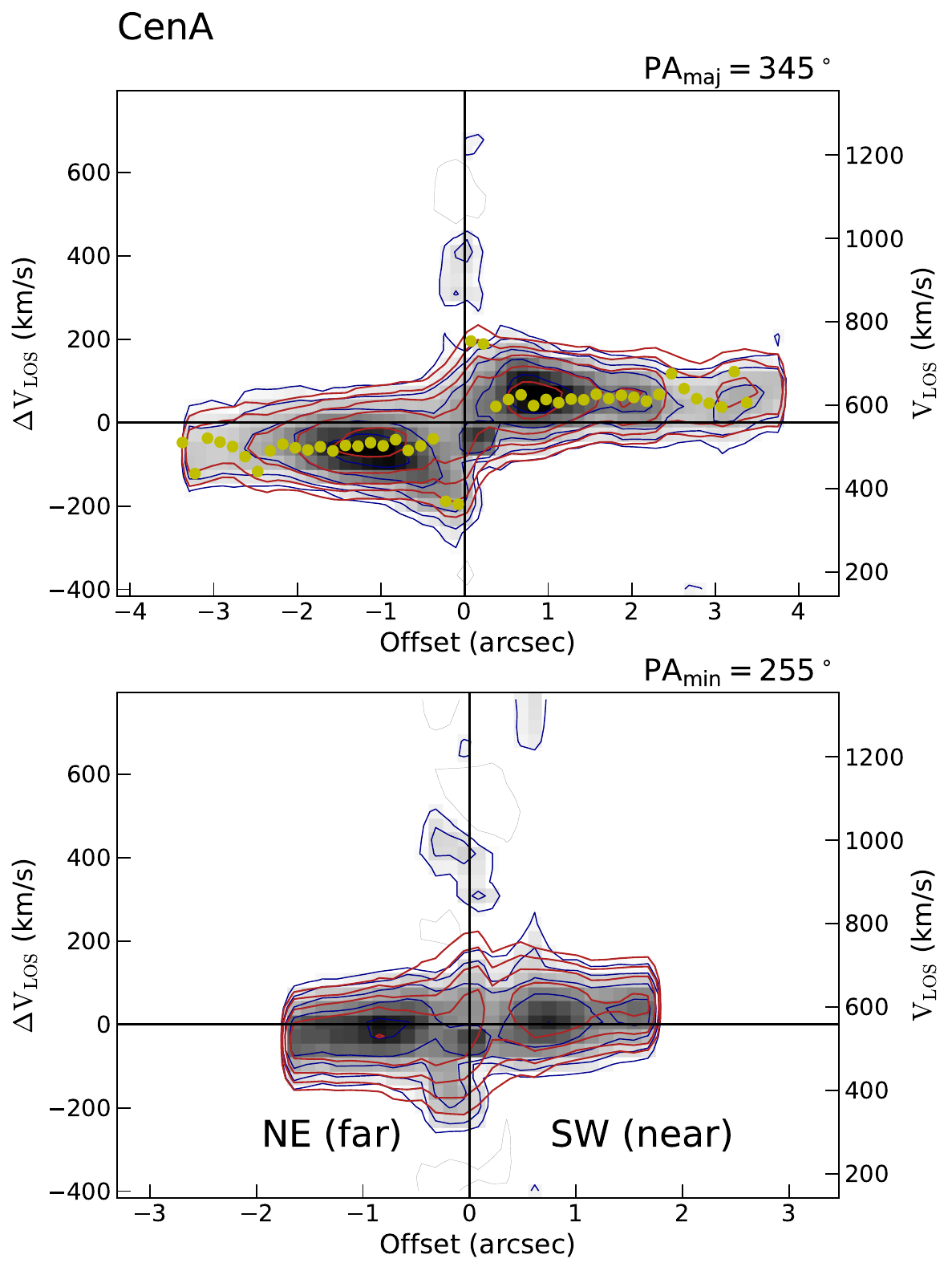}
\caption{Same as Fig.~\ref{fig:BAROLOpv_H2S5NOVRAD} but for a  $^{\rm
    3D}$BAROLO model with a  $v_{\rm RAD}$ component.}
     \label{fig:BAROLOpv_H2S5}
\end{figure}

\subsection{Modeling of the gas kinematics}\label{subsec:kinematicmodeling}
To model the gas kinematics we used the $^{\rm 3D}$BAROLO code
\citep{DiTeodoro2015} 
to fit line emission data cubes with disk
models with 3D tilted 
rings, which is needed owing to the presence of a warped gaseous dusty
disk on different physical scales \citep[see e.g.,][]{Quillen1992,
    Quillen2006, Eckart1999, Neumayer2007, Espada2009}.
This code has already been successfully applied to 
MRS observations of the nearby Seyfert galaxy MCG–05-23-16 \citep[see
e.g.,][]{EsparzaArredondo2025}. In this 
section, we focus on
the molecular gas H$_2$ S(5) and S(1) transitions and  the ionized
gas lines [Ar\,{\sc ii}] and [Ne\,{\sc iii}]. The free parameters of
a $^{\rm 3D}$BAROLO  disk model are: the kinematic
center,  $v_{\rm sys}$, the disk inclination ($i$), PA of the major
axis (${\rm   PA}_{\rm maj}$), the
scale height of the disk, the circular  velocity ($v_{\rm rot}$), and the velocity
dispersion ($\sigma_{\rm gas}$). It also accounts for
effects related to the beam smearing 
in the definition of the best-fit model for a given spatial
resolution. For each MRS data set, we used  the measured 
spatial FWHM of the smoothed data cubes 
(Sect.~\ref{subsec:analysis}) as the beam value.  For this analysis,
we used the continuum subtracted data cubes around the lines produced with the
{\sc Alucine} tool (see Sect.~\ref{subsec:analysis}).

The $^{\rm 3D}$BAROLO code allows the user to fix any of the above parameters. For the
kinematic center we took the position of the continuum peak and fixed the scale
height of the disk to 0.5\arcsec (20\,pc). The latter is within
the range estimated by \cite{Espada2017} for Cen~A on similar physical
scales.  The minimum value of
$\sigma_{\rm gas}$ was set to the instrumental resolution at the
line wavelength. After the first run, we fixed $v_{\rm
  sys}$ and fit ${\rm   PA}_{\rm maj}$, $i$, $v_{\rm rot}$, and
$\sigma_{\rm gas}$ again.
This code also generates maps
of the velocity-integrated intensity (moment 0), the 
mean-velocity field (moment 1), and
velocity dispersion field (moment 2) of the observations and best-fit model, as well as 
position-velocity (p-v) diagrams along the major and minor kinematic
axes.

\begin{figure}
\hspace{0.25cm}
  \includegraphics[width=8.5cm]{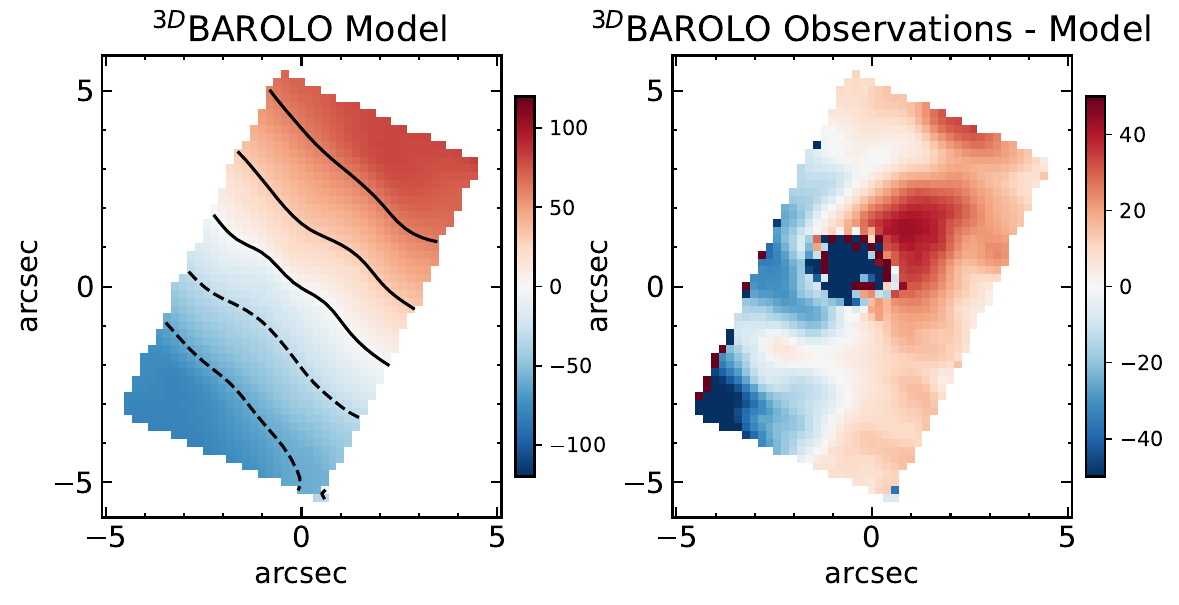}

\hspace{0.25cm}
  \includegraphics[width=8.5cm]{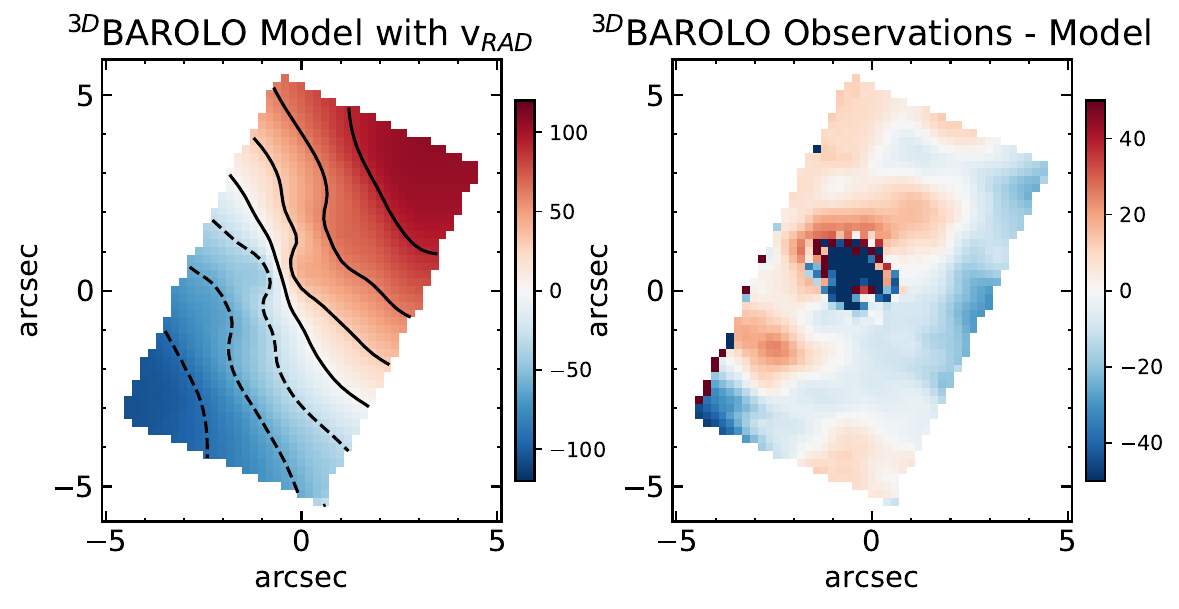}
  \caption{Same as Fig.~\ref{fig:BAROLOvfields_H2S5} but for the
    velocity fields of H$_2$ S(1). Orientation, color scale, and contours for the velocity models are as in
     Fig.~\ref{fig:maintextalucinemapsH2}.} 
     \label{fig:BAROLOvfields_H2S1}
\end{figure}

\subsubsection{Results for H$_2$ S(5) and H$_2$ S(1)}
We started by modeling the H$_2$ S(5) and H$_2$ S(1) transitions because
these generally present smaller deviations from rotation than
ionized gas lines \citep[see e.g.,][]{Davies2024}. For the initial
estimates of the PA$_{\rm maj}$ and $i$ we took values comprised within those compiled by
\cite{Quillen2010} in the radial distance range of our observations
and let
them to vary by $\pm 25^{\rm o}$. The H$_2$ S(5) model mean velocity field and
residuals computed by subtracting it from the $^{\rm 3D}$BAROLO
moment 1 map are in Fig.~\ref{fig:BAROLOvfields_H2S5}, left and
right top panels, respectively. The latter reveals some deviations from rotational motions,
reminiscent of an S-shape centered at the AGN position and
more prominent deviations to the northwest (NW). These may be related
to  gas streamers or filaments  detected in cold molecular gas, which
connect the CND,
which is beyond our MRS observations, and the nuclear ring \citep[see Fig.~4
of][]{Espada2017}. We note that some of the distorsions of the
observed velocity field are accounted for with the fit model since
there are some variations of the kinematical PA$_{\rm maj}$
and $i$ (Fig.~\ref{fig:BAROLOparameters_H2S5NOVRAD}). These are
similar to those of the warped disk
model used to explain the kinematics of  the hot
molecular gas  on comparable scales
\citep[see][]{Neumayer2007, Quillen2010}.

\begin{figure}
\hspace{0.25cm}
  \includegraphics[width=8.5cm]{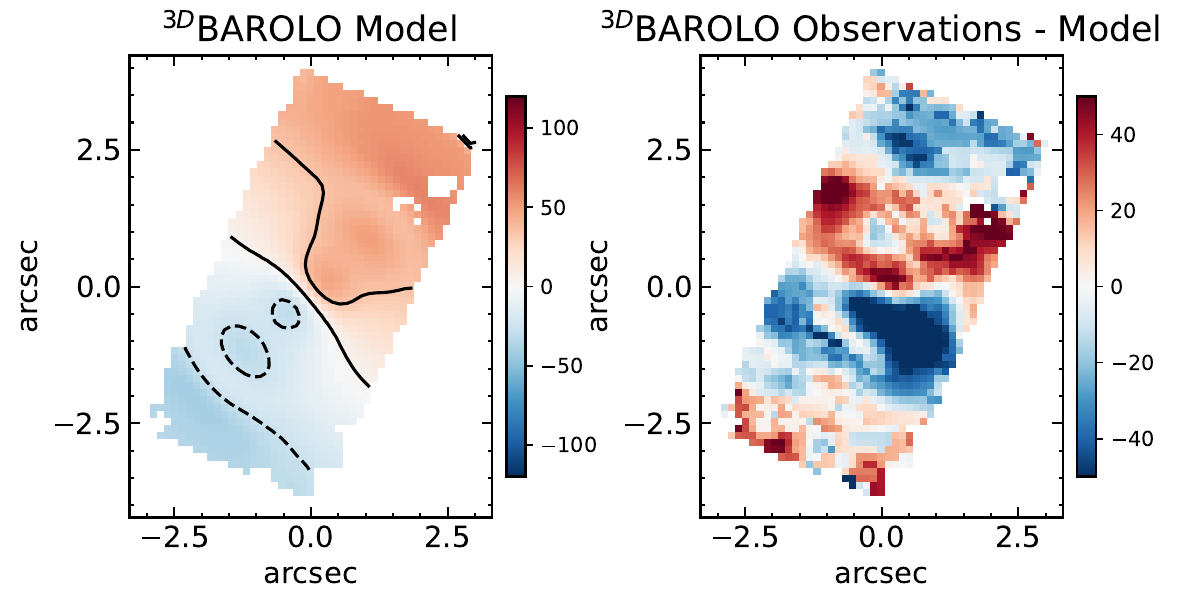}
  \caption{Same as  Fig.~\ref{fig:BAROLOvfields_H2S5} (top)
but for the velocity fields of [Ar\,{\sc ii}]. Orientation, color
scale, and contours for the velocity models as in 
     Fig.~\ref{fig:maintextalucinemapsH2}.}
     \label{fig:BAROLOvfields_ArII}
\end{figure}

The p-v diagrams along the mean fit kinematic major and minor axes of the nuclear
H$_2$ S(5) disk show emission spreading over a
range of velocities around $v_{\rm sys}$ of [-150, +180]$\,{\rm km
  \,s}^{-1}$ (see Fig.~\ref{fig:BAROLOpv_H2S5NOVRAD}). It is
noticeable that the minor axis p-v diagram does not present
strong deviations from rotation. Within the central $\pm
0.5\arcsec$, some of the observed velocity structure, which is 
not reproduced by the model, might be artefacts left from the
subtraction of the strong continuum associated with the AGN.

To account for the velocity deviations seen between  model
and observations, we fit a new $^{\rm 3D}$BAROLO model including a
radial velocity ($v_{\rm RAD}$)
component. We allowed this parameter as well as  PA$_{\rm maj}$,
$i$, $v_{\rm rot}$, and $\sigma_{\rm gas}$ to 
vary. We note that a scenario with radial motions over a pure warping of the disk was favored
by \cite{Espada2017}. The resulting new velocity model and corresponding residuals are
presented in 
Fig.~\ref{fig:BAROLOvfields_H2S5} (lower panels).  Assuming that the galaxy near
side in the inner region of Cen~A is to the southwest 
\citep[SW, see][]{Quillen2006} and the modeled radial motions are in the disk of the
galaxy, this would imply streaming (inflows) motions 
toward the center of the galaxy, as were suggested by the S-shape seen
in the velocity dispersion map (Fig.~\ref{fig:maintextalucinemapsH2}). The residual velocities
are slightly smaller when compared with the warped model
with no $v_{\rm RAD}$, except along the minor axis of the
galaxy. Nevertheless, the $v_{\rm RAD}$ model reproduces fairly well the emission seen along the
minor axis p-v diagram  (see Fig.~\ref{fig:BAROLOpv_H2S5}, bottom).
It is possible that part of the nuclear noncircular motions are also
taking place  outside the plane of the galaxy.  However, these are not
strong, given the weak blueshifted components seen in 
the nuclear H$_2$ line profiles (Fig.~\ref{fig:lineprofiles_H2S5}). 

We performed the same analysis for the H$_2$ S(1) line. The
 residuals left after subtracting $^{\rm
  3D}$BAROLO models without and with a $v_{\rm RAD}$ component
(Fig.~\ref{fig:BAROLOvfields_H2S1}) show that the latter appears to reproduce better 
the S-like shape distorsion to the NW of the AGN. However, this also
produces more blueshifted residuals to the SW. The p-v
diagrams (Fig.~\ref{fig:BAROLOpv_H2S1NOVRAD}) are strongly affected by
the imperfect continuum subtraction in the inner $2\arcsec$. Further
away, the model without $v_{\rm RAD}$ fits the circular
motions along the major axis of the galaxy relatively well, but deviations are present
to the southeast of the AGN.

In summary, the warm molecular gas kinematics reveal rotation but they are
strongly affected by complex noncircular
motions. Some can be fit either with a nuclear warped disk or by
including a $v_{\rm RAD}$ component (or a combination of both), but
some velocity residuals still
remain.  The radial motions are likely associated
with gas streamers towards the center of the galaxy seen in the cold
molecular gas \citep{Espada2017}.

\begin{figure}
  \hspace{1cm}
  \includegraphics[width=7cm]{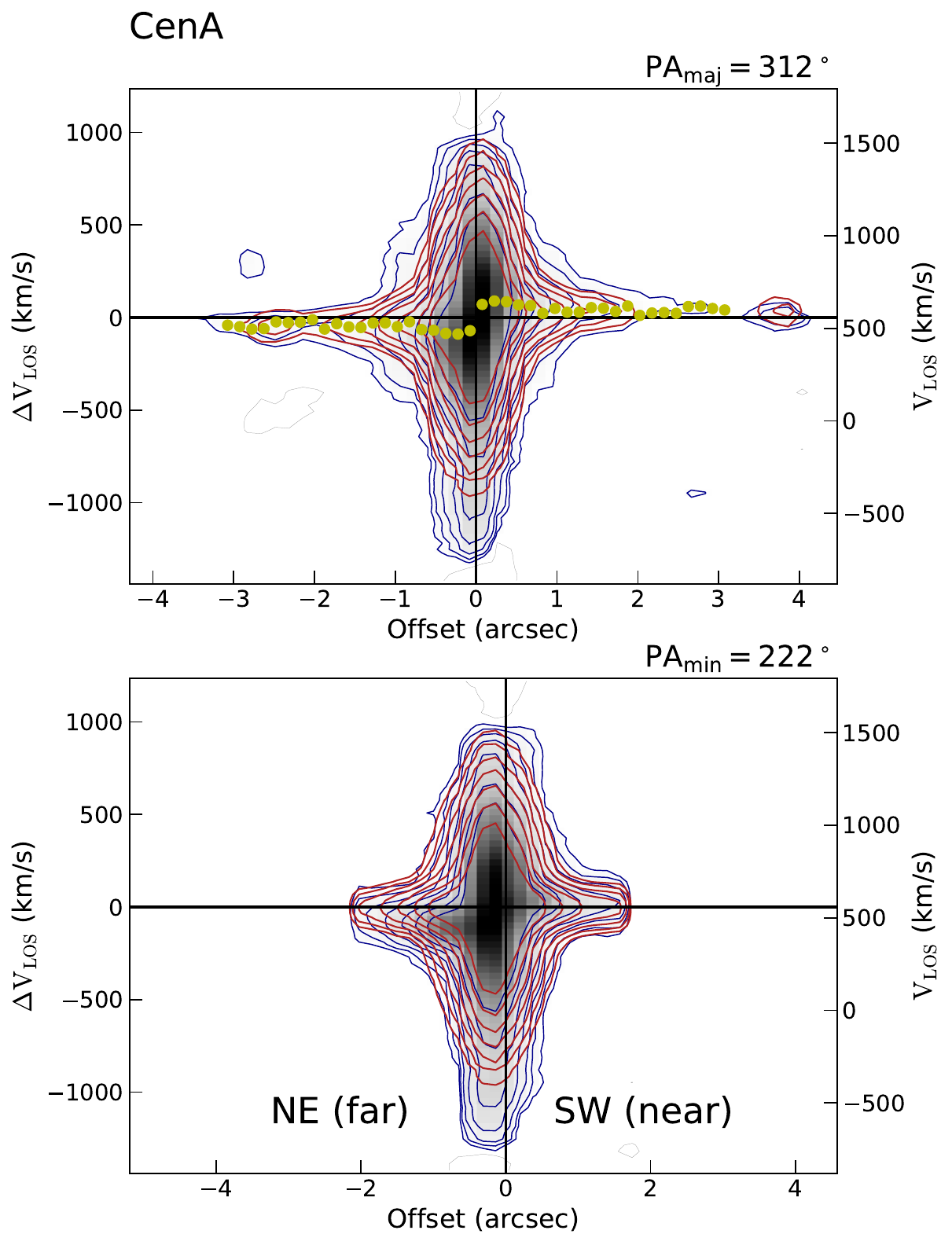}
\caption{P-v diagrams extracted along the kinematic major axis (top)
  and minor axis (bottom) for [Ar\,{\sc ii}]. The disk 
  model does not include a $v_{\rm RAD}$ component. Colors and
  contours as in Fig.~\ref{fig:BAROLOpv_H2S5NOVRAD}.}
     \label{fig:BAROLOpv_ArIINOVRAD}
\end{figure}

\subsubsection{Results for [Ar\,{\sc ii}] and [Ne\,{\sc iii}]}\label{subsubsec:BAROLOionizedgas}
In this section we fit $^{\rm
  3D}$BAROLO models to the bright ionized gas emission lines [Ar\,{\sc
  ii}] and [Ne\,{\sc iii}]. As for the molecular gas, we fixed the kinematic center,
v$_{\rm sys}$, and in this case, PA$_{\rm maj}$ and $i$  were only
allowed  to vary within $\pm 10^\circ$ of the average values 
obtained for the H$_2$ S(5) and S(1) line fits without $v_{\rm RAD}$
(Fig.~\ref{fig:BAROLOparameters_H2S5NOVRAD}). We fit  these two
parameters as well as the $\sigma_{\rm gas}$ and $v_{\rm rot}$. 

The residual velocity map after subtracting the fit $^{\rm
  3D}$BAROLO model reveals
velocities  with strong deviations from a purely rotating disk (see
Fig.~\ref{fig:BAROLOvfields_ArII} for [Ar\,{\sc ii}] and Fig.~\ref{fig:BAROLOvfields_NeIII}
for [Ne\,{\sc iii}]). The velocity residuals resemble the mean-velocity
fields of the high IP lines [Ne\,{\sc vi}] and [Ne\,{\sc v}]
(see Figs.~\ref{fig:maintextalucinemapsFSL} and \ref{fig:appendixalucinemaps}), which
likely have a very low rotational component. These are readily
understood by inspecting the p-v diagrams 
(Figs.~\ref{fig:BAROLOpv_ArIINOVRAD} and
\ref{fig:BAROLOpv_NeIIINOVRAD}) extracted along the 
kinematic major and minor axes. As already inferred from the nuclear
profiles of the two lines (Figs.~\ref{fig:lineprofiles_ArII} and \ref{fig:lineprofiles_extra}), at the
AGN position the ionized gas  presents fast velocities reaching
approximately $+1000\,{\rm km\,s}^{-1}$ and
$-1400\,{\rm km\,s}^{-1}$ along both kinematic axes. These are similar
to values detected in the ionized gas of IC~5063 \citep[see Fig.~10 of ][]{Dasyra2024}. 
The fast ionized gas velocities in Cen~A 
are associated with the unresolved second component derived from our 
spaxel-by-spaxel analysis of the [Ar\,{\sc ii}] line, which 
can be seen in the central 2\arcsec, approximately (bottom panels of
Fig.~\ref{fig:appendixalucinemapstwocomponents}).

At the AGN position,
the $^{\rm 3D}$BAROLO model for [Ar\,{\sc ii}] reproduces velocities of up 
to  $\sim \pm 900\,{\rm km\,s}^{-1}$ by fitting values of $\sigma_{\rm
  gas} \sim 300\,
\,{\rm km\,s}^{-1}$.
At approximate radial 
distances of 1-2\arcsec, and for both lines, there are further deviations from the model
circular motions, both blueshifted and redshifted, with velocities of
up to $500\,{\rm km\,s}^{-1}$ in the p-v diagrams. This indicates that
noncircular motions are probably both in the plane of the 
galaxy and outside. Finally, along the
kinematic major axis at radial distances greater than
$\simeq 2\arcsec$, the gas is rotating with low velocity dispersions
compared with the nuclear region.

\cite{Neumayer2007} identified non-rotational motions in the central
1\arcsec \, of Cen~A with
the near-IR line  [Si\,{\sc vi}], which appeared to be more prominent
to the SW of the AGN.  
Indeed, our [Ne\,{\sc vi}] mean-velocity map
(Fig.~\ref{fig:maintextalucinemapsFSL}, second panel from the top) is
virtually identical to  the coronal line  [Si\,{\sc
    vi}] velocity field (their figure~3). These authors interpreted the blueshifted component as a
backflow of gas which was  
accelerated by the radio jet.  Furthermore, they pointed out that
such behavior is predicted by jet simulations on larger physical
scales. From these new MRS observations, we conclude that the majority of
noncircular motions of several hundreds of kilometers per second in the
ionized gas, which are observed  in several directions
within the inner region, are likely indicating gas outflowing both in the disk
and outside the plane of the galaxy. They resemble the motions of an
expanding bubble, possibly
induced by the passage of the
intermediate-power radio jet of Cen~A through the galaxy ISM
\citep[see e.g.,][]{Mukherjee2016}. Although on very different
  physical scales, a similar scenario was proposed for the 
  luminous quasar 3C273 \citep{Husemann2019}.

\subsection{Energetics of the nuclear ionized gas outflow}\label{subsec:outflowenergetics}
In  previous sections we established the
presence of a nuclear fast ionized gas outflow.  To estimate its properties we
follow the methodology applied to MRS observations by \cite{HermosaMunoz2025}. Briefly, we
first derived the mass of ionized gas ($M_{\rm out}$) in the nuclear
region using
the Pf$\alpha$ line (Table~\ref{tab:nuclearfluxes}) and their equation~(1). We estimated an electron
density of $n_{\rm e}=868\,{\rm cm}^{-3}$ using the ratio between the two
[Ne\,{\sc v}] lines and {\sc pyneb} \citep{Luridiana2015}. Using the nuclear Pf$\alpha$ flux
(Table~\ref{tab:nuclearfluxes}), the mass of ionized gas in the nuclear
(unresolved) region of Cen~A is $M_{\rm ion} \simeq
10^4\,M_\odot$. The Pf$\alpha$ line is broad and relatively faint against the
bright continuum (see
Fig.~\ref{fig:nuclearspectrum_channels}),  which complicates an
accurate kinematic
decomposition. To estimate the ionized gas mass in the outflow we
assumed that, similarly to [Ar\,{\sc ii}] and 
[Ne\,{\sc iii}], approximately half of the flux is in the broad
component as derived from the fits to the lines with two Gaussians. 
Thus the outflowing ionized gas mass is $M_{\rm out}=5 \times
10^3\,M_\odot$, where the uncertainties include the Pf$\alpha$ flux
error ($10\%$) and the fraction of the line flux associated with the
fast velocities. For the maximum outflow velocity, 
 we took the expression $v_{\rm out-max} = \Delta v+2 \times
 \sigma_{\rm broad}$, where $\Delta v$
 is the shift (in absolute value) of the line with respect to $v_{\rm
   sys}$. Relying again on [Ar\,{\sc ii}] and [Ne\,{\sc iii}], we
 obtained estimates for the maximum outflow velocity 
 of $v_{\rm out-max}=1183$ and $652\,{\rm km\,s}^{-1}$, respectively. The lower
 value from [Ne\,{\sc iii}] comes, at least in part, from the fact
 that the two Gaussian fit does not account for the high
 velocity line wings  (see middle panel of Fig.~\ref{fig:lineprofiles_extra}).

 Using $dM_{\rm out}/dt =
 3\times  M_{\rm out} \times v_{\rm out-max}/R_{\rm out}$
\citep[][and references therein]{HermosaMunoz2025} and  the region
radius ($R_{\rm 
  out}$) that of  the nuclear source,
we derived an ionized gas mass outflow rate of $dM_{\rm
  out}/dt=1.6-2.9\,M_\odot \,{\rm yr}^{-1}$. The quoted values take
into account the different estimates for the outflow maximum
velocity. 
This relatively large value
results in part from the fact that the nuclear fast ionized gas outflow in
Cen~A stems from an extremely  compact region. We note that the
nuclear region is unresolved, and thus this 
mass outflow rate is a lower limit. 

In Sects.~\ref{subsec:nuclearlineratios} and
\ref{subsec:extendedemission} we presented evidence that both an AGN and
shocks may contribute to the excitation of the gas. Moreover, the complex
ionized and molecular gas kinematics suggest a strong interaction
between the radio jet and the galaxy ISM on the nuclear scales of
Cen~A. The resulting ionized gas outflow might be driven by the jet and/or the
AGN. As done for NGC~1068 by \cite{GarciaBurillo2014}, we can evaluate
whether both mechanisms can inject sufficient energy to launch the
ionized gas  outflow. The maximum kinetic energy of the outflow ($L_{\rm kin}=1/2 \times
dM_{\rm out}/dt \times (v_{\rm out})^2$) is approximately
$1.3 \times 10^{42}\,{\rm erg\,s}^{-1}$, and thus the estimated  jet power
of $P_{\rm jet} \sim 2\times 10^{43}\,{\rm erg \,
  s}^{-1}$  (see Sect.~\ref{sec:introduction}) would be sufficient. On the other hand,
the momentum provided by AGN photons is generally computed as
$L_{\rm bol}{\rm (AGN)}/c$. Thus $L_{\rm bol}{\rm (AGN)}$ gives a
momentum range of approximately $3\times10^{32}-1\times 10^{33}\,{\rm g\,cm\,s}^{-2}$,
while the maximum outflow momentum flux ($dP_{\rm out}/dt= dM_{\rm out}/dt \times
v_{\rm out}$) is of the order of $2\times 10^{34}\,{\rm
  g\,cm\,s}^{-2}$. The momentum boost needed would be  approximately
16-64, which is a reasonable factor according to numerical
simulations \citep[see][]{FaucherGiguere2012,
  GarciaBurillo2014}. However, based on this analysis,  we
cannot determine which is the dominant mechanism. Nevertheless,  Cen~A's radio jet is traveling at
subluminal apparent velocities \citep[$v\sim
0.5c$,][]{Hardcastle2003}. Thus it is likely that it is the main influence on
these scales. 

\section{Discussion and summary}\label{sec:discussion}

We have analyzed  MIRI-MRS $5-28\,\mu$m observations of the inner $\simeq
7-14\arcsec$ ($\simeq 100-200\,$pc) region of
Cen~A which we used to investigate in detail the kinematics of both
the ionized and the warm molecular gas. This work is
part of the MIRI GTO program MICONIC. 

Morphology of the gas: The two gas phases explored with the
  MIRI-MRS observations
present remarkably different morphologies  in the central
  100-200\,pc region. The warm
molecular gas traced with the H$_2$ S(1) transition  shows a ring-like
shape with a more pronounced
gas deficit at the AGN position  than  the S(5) line. The latter
  line probes  warmer
gas further inside the nuclear disk 
  (Fig.~\ref{fig:maintextalucinemapsH2}). The fine structure lines
show the 
brightest emission along the direction of the radio jet,
extending over the $\sim 100$\,pc  region mapped with MRS
(Figs.~\ref{fig:maintextalucinemapsFSL} and  
\ref{fig:appendixalucinemaps}). This structure constitutes the base of
the much larger ionization  
cone identified with {\it Spitzer}/IRS mapping observations of  [O\,{\sc iv}] and [Ne\,{\sc v}] 
\citep{Quillen2008}.  Emission from other lines (e.g., [Ar\,{\sc ii}], [Ne\,{\sc
  ii}], [Ne\,{\sc iii}], and [S\,{\sc iii}]), on the
  other hand, is also present  in the disk of the galaxy.

Kinematics of the ionized gas: The fine-structure
  lines, from low IP ([Fe\,{\sc ii}] and [Ar\,{\sc
  ii}]) to high IP ([Ne\,{\sc vi}]), as well as the warm
molecular gas lines, show  extremely complex kinematics. There is evidence for rotation, 
seen in both gas phases, from modeling the observations with tilted
rotating rings using the $^{\rm 3D}$BAROLO tool
\citep{DiTeodoro2015}. More importantly, we detected several ionized
gas kinematical features that agree with  
predictions from simulations of the passage of low-intermediate 
power radio jets \citep{Mukherjee2016, Mukherjee2018, Talbot2022,
  Meenakshi2022, Borodina2025} through the ISM of a galaxy. These include:

\begin{itemize}

\item
  Fast velocities in the ionized gas are seen as broad line components
in the nuclear profiles of [Ar\,{\sc ii}]  and [Ne\,{\sc
  iii}] with $\sigma \simeq 600\,{\rm km\,s}^{-1}$
  (Figs.~\ref{fig:lineprofiles_ArII} and \ref{fig:lineprofiles_extra})
  and in p-v diagrams
with values of up to approximately $+1000\,$km\,s$^{-1}$ and $-1400\,$km\,s$^{-1}$
(Figs.~\ref{fig:BAROLOpv_ArIINOVRAD} and
\ref{fig:BAROLOpv_NeIIINOVRAD}). These appear to be confined within the 
unresolved nuclear region (size of $\simeq$0.35\arcsec = 6\,pc, FWHM) of Cen~A. Velocities of
hundreds of kilometers per second  are also observed on slightly larger scales
and in several directions,
up to projected radial
distances of $\simeq 2\arcsec$=35\,pc.
This is a clear observational signature
of the presence of a jet-inflated bubble/outflow in the central region. 
Further  out there are (perturbed) rotational motions in the disk of
the galaxy, which are identified in the low IP emission lines (i.e.,
[Ar\,{\sc ii}] and [Ne\,{\sc iii}]).

\item
  Velocity dispersion enhancements are observed in all the
fine-structure lines, both at the AGN position 
as well as extended over a few arcseconds in the direction perpendicular
to the radio jet (Figs.~\ref{fig:maintextalucinemapsFSL} and 
\ref{fig:appendixalucinemaps}, right panels). In the radio jet simulations
above these are  explained by  the lateral
expansion of the jet as it interacts with the galaxy
ISM. Observationally these have been  found in other  
Seyfert galaxies with jets \citep{Venturi2021, PeraltadeArriba2023, Davies2024,
  Zhang2024}. 

\item
  The nuclear [Ne\,{\sc v}]/[Ne\,{\sc ii}], [Ne\,{\sc iii}]/[Ne\,{\sc
  ii}], and [Fe\,{\sc ii}]/Pf$\alpha$ ratios of Cen~A are intermediate
between those of Seyfert and type 2 QSO nuclei and those of LLAGNs and
shocked regions
(Figs.~\ref{fig:diagramratios} and
\ref{fig:diagramratios_feiitopfalpha}). Comparison of neon line ratios
with fast shock models with  
and without a small contribution from AGN
photoionization \citep{Alarie2019, Feltre2023} shows that shocks are
likely to play an important role in the excitation of 
the  ionized gas of Cen~A's  nuclear region. Interestingly, in
regions along the  radio jet direction, just outside 
the very nuclear region, the neon line ratios become higher
(Fig.~\ref{fig:lineratiomaps}), suggesting  that the gas is also being
 illuminated by the AGN. 
\end{itemize} 

We derived an ionized gas mass outflow rate in the nuclear region
(central 6\,pc) of
$1.6-2.9\,M_\odot \,{\rm yr}^{-1}$. Based on estimates for $L_{\rm bol}{\rm (AGN)}$
 \citep{Israel1998, Beckmann2011} and $P_{\rm jet}$ 
\citep{Croston2009, Wykes2013, Neff2015}, we concluded that both mechanisms can inject sufficient
energy to launch the outflow.

Kinematics of the warm molecular gas: The mid-IR H$_2$ line
kinematics in the central region of Cen~A are reproduced 
with  a warped rotating disk model and/or  a rotating
  disk with radial motions
(Figs.~\ref{fig:BAROLOvfields_H2S5} and \ref{fig:BAROLOvfields_H2S1},
top and bottom panels, respectively). The latter could be
associated with gas streamers detected in cold molecular gas
\citep{Espada2017}, although some non-rotational motions could also
take place outside the plane of  the galaxy. Notably, there is no strong
evidence for a fast nuclear molecular gas outflow in Cen~A, only a
weak blueshifted wing (see the H$_2$ S(5) line profile in
Fig.~\ref{fig:lineprofiles_H2S5}). This is 
unlike the nuclear warm molecular gas outflow detected in IC~5063 \citep{Dasyra2024}, although on much 
larger scales, which results from the radio jet  crossing the galaxy disk. 
Several factors could  influence the central region of Cen~A.
There is a deficit of molecular gas in that region, including the cold phase \citep[see 
also][]{Espada2017}. Additionally, Cen~A's jet is launched perpendicular to the
galaxy disk, thus limiting the geometrical coupling, although
simulations do predict some interaction with the galaxy ISM. Indeed,  molecular outflows driven by a radio
jet perpendicular to the disk have been observed, for  example, in the 
 Seyfert galaxy NGC~613 \citep{Audibert2019}. In contrast to Cen~A,
that galaxy has a high nuclear
molecular gas concentration, both hot and cold, when compared to inner $\sim 200\,$pc region
\citep{GarciaBurillo2021, GarciaBurillo2024}.

In summary, our new MIRI-MRS observations of Cen~A provide clear
evidence of the presence of an ionized gas 
outflow whose fastest velocities ($|v|>$500$\,{\rm km\,s}^{-1}$)
originate from the central $6\,$pc. The noncircular motions
and other properties of  the ionized gas
are consistent with an expanding bubble, likely induced by the passage
of the intermediate-power radio jet through the galaxy
ISM. The absence of an associated fast molecular gas outflow is probably
due to a combination of a lack of large warm H$_2$ column densities in the nuclear region
and the fact that the inner radio jet appears to be perpendicular to 
the galaxy CND \citep{Israel1998}. This may limit
the  coupling of the radio jet with the ISM  on the physical scaled mapped with MIRI-MRS.

    \begin{acknowledgements}

 We thank the referee for a constructive report. We are grateful
to Cristina Ramos Almeida and Lulu Zhang for sharing the model 
curves plotted in Fig.~\ref{fig:diagramratios}. AAH and LHM acknowledge
support from grant PID2021-124665NB-I00 funded by the Spanish
Ministry of Science and Innovation and the State Agency of Research
MCIN/AEI/10.13039/501100011033  and ERDF A way of making Europe. LC acknowledges support by
grant PIB2021-127718NB-100 from the Spanish Ministry of Science 
and Innovation/State Agency of Research MCIN/AEI/10.13039/501100011033
and by “ERDF A way of making Europe”.  G\"O
acknowledges support from the Swedish 
National Space Agency (SNSA). SGB acknowledges support from the
Spanish grant PID2022-138560NB-I00, funded by
MCIN/AEI/10.13039/501100011033/FEDER, EU.

MIRI draws on the scientific and technical expertise of the following organisations:
Ames Research Center, USA; Airbus Defence and Space, UK; CEA-Irfu, Saclay,
France; Centre Spatial de Liége, Belgium; Consejo Superior de Investigaciones
Científicas, Spain; Carl Zeiss Optronics, Germany; Chalmers University of Technology, Sweden; Danish Space Research Institute, Denmark; Dublin Institute
for Advanced Studies, Ireland; European Space Agency, Netherlands; ETCA,
Belgium; ETH Zurich, Switzerland; Goddard Space Flight Center, USA; Institute d’Astrophysique Spatiale, France; Instituto Nacional de Técnica Aeroespacial, Spain; Institute for Astronomy, Edinburgh, UK; Jet Propulsion Laboratory,
USA; Laboratoire d’Astrophysique de Marseille (LAM), France; Leiden University, Netherlands; Lockheed Advanced Technology Center (USA); NOVA OptIR group at Dwingeloo, Netherlands; Northrop Grumman, USA; Max-Planck
Institut für Astronomie (MPIA), Heidelberg, Germany; Laboratoire d’Etudes
Spatiales et d’Instrumentation en Astrophysique (LESIA), France; Paul Scherrer Institut, Switzerland; Raytheon Vision Systems, USA; RUAG Aerospace,
Switzerland; Rutherford Appleton Laboratory (RAL Space), UK; Space Telescope Science Institute, USA; Toegepast- Natuurwetenschappelijk Onderzoek
(TNO-TPD), Netherlands; UK Astronomy Technology Centre, UK; University
College London, UK; University of Amsterdam, Netherlands; University of
Arizona, USA; University of Cardiff, UK; University of Cologne, Germany;
University of Ghent; University of Groningen, Netherlands; University of Leicester, UK; University of Leuven, Belgium; University of Stockholm, Sweden; Utah State University, USA. A portion of this work was carried out at
the Jet Propulsion Laboratory, California Institute of Technology, under a contract with the National Aeronautics and Space Administration. We would like
to thank the following National and International Funding Agencies for their
support of the MIRI development: NASA; ESA; Belgian Science Policy
Office; Centre Nationale D’Etudes Spatiales (CNES); Danish National Space Centre;
Deutsches Zentrum fur Luft-und Raumfahrt (DLR); Enterprise Ireland; Ministerio De Economía y Competitividad; Netherlands Research School for Astronomy (NOVA); Netherlands Organisation for Scientific Research (NWO);
Science and Technology Facilities Council; Swiss Space Office; Swedish National Space Board; UK Space Agency. This work is based on observations
made with the NASA/ESA/CSA James Webb Space Telescope. The data were
obtained from the Mikulski Archive for Space Telescopes at the Space Telescope Science Institute, which is operated by the Association of Universities for
Research in Astronomy, Inc., under NASA contract NAS 5-03127 for JWST;
and from the European JWST archive (eJWST) operated by the ESDC.

This research has made use of the NASA/IPAC Extragalactic Database (NED),
which is operated by the Jet Propulsion Laboratory, California Institute of Technology,
under contract with the National Aeronautics and Space
Administration.

This research made use of NumPy \citep{Harris2020}, Matplotlib
\citep{Hunter2007} and Astropy \citep{Astropy2013, Astropy2018}.

\end{acknowledgements}

   \bibliographystyle{aa} 
   \bibliography{bibliography} 

   \begin{appendix}
   \onecolumn  

  \section{Additional figures for spectra, line profile fit,  {\sc
      alucine} maps, and  $^{\rm 3D}$BAROLO analysis}\label{appendix:linemaps}
\begin{figure*}[hbt!]
\hspace{2cm}
     \includegraphics[width=14cm]{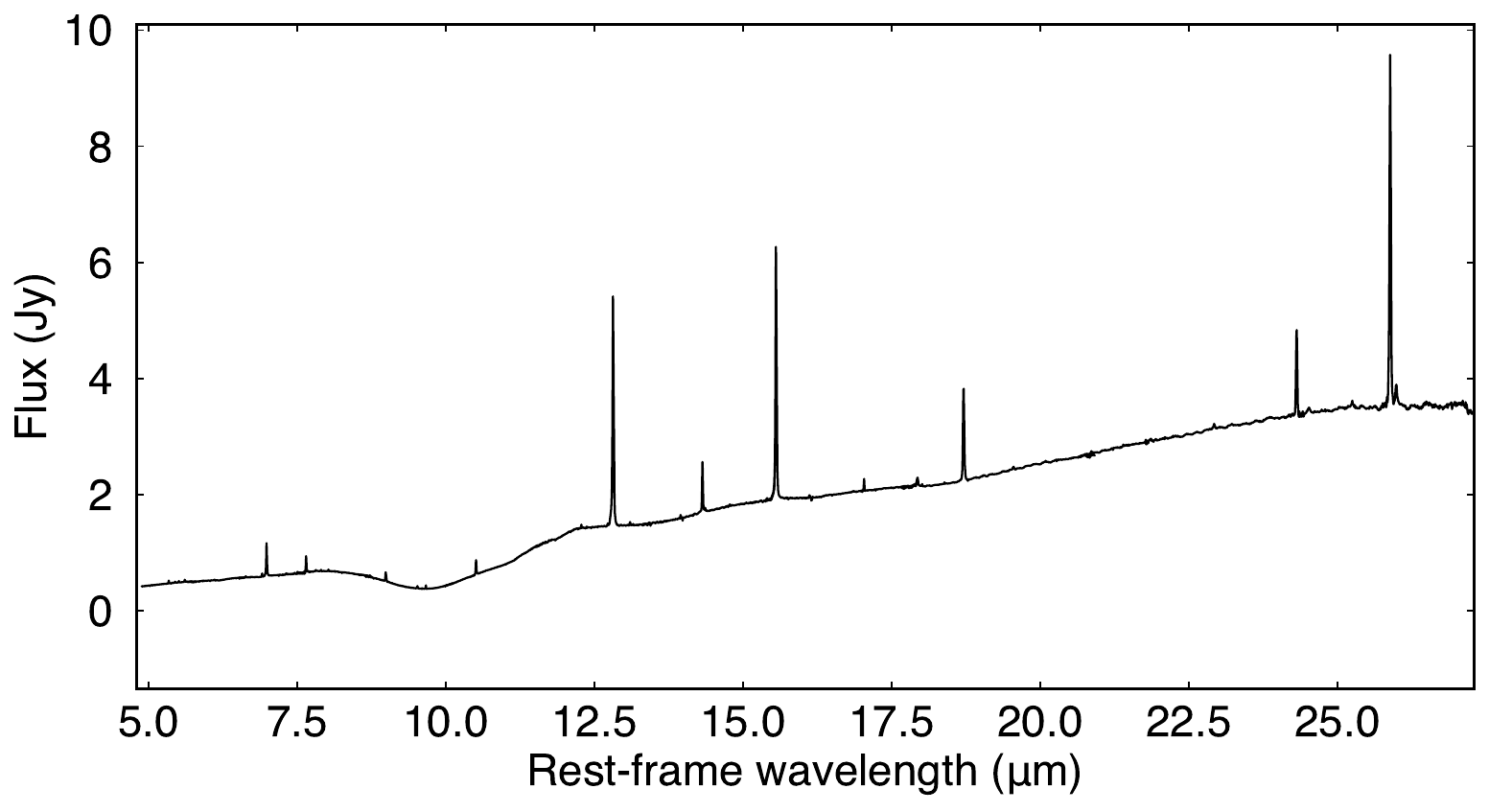}
     \caption{Full $5-27.3\,\mu$m spectrum of
  Cen~A extracted as a point source. Those spectra separated by individual channels are in
  Fig.~\ref{fig:nuclearspectrum_channels}.} 
     \label{fig:fullspectrum}
\end{figure*}

\begin{figure*}[hbt!]
    \hspace{3cm}
     \includegraphics[width=12cm]{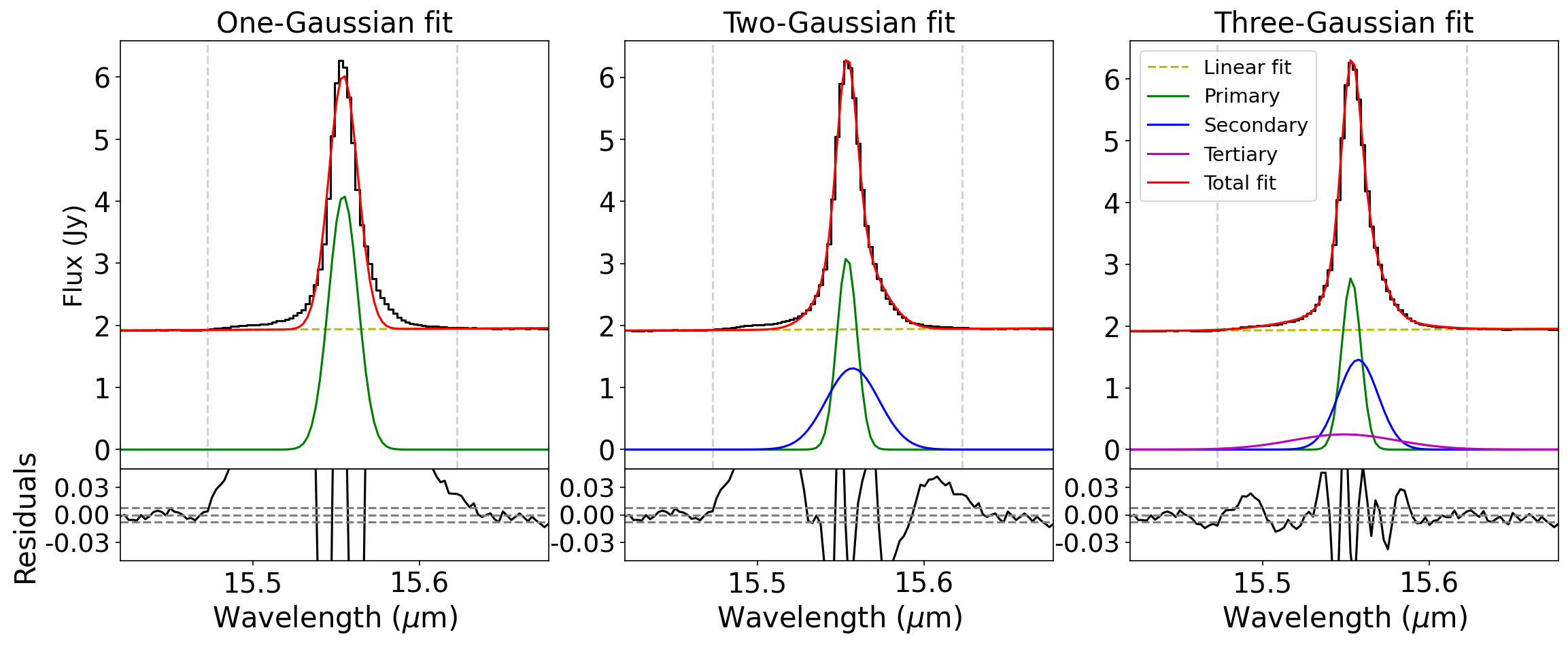}
     \caption{Same as Fig.~\ref{fig:lineprofiles_ArII} but for the
parametric fits of the line profile of
       [Ne\,{\sc iii}].}
     \label{fig:lineprofiles_extra}
\end{figure*}

  In this appendix we show the full $5-27\,\mu$m nuclear spectrum of
  Cen~A in Fig.~\ref{fig:fullspectrum} and the parametric fit to the nuclear [Ne\,{\sc
  iii}] line in Fig.~\ref{fig:lineprofiles_extra} with one, two, and
three Gaussians. Table~\ref{tab:parametricfits} summarizes the
results for the fit peak velocities and velocity dispersions of the
components for this line as well as for H$_2$ S(5) and
[Ar\,{\sc ii}] (fits plotted in
Figs. ~\ref{fig:lineprofiles_H2S5} and ~\ref{fig:lineprofiles_ArII},
respectively).

\begin{figure*}[hbt!]

  \vspace{-0.35cm}
 \hspace{1.5cm} 
  \includegraphics[width=14.7cm]{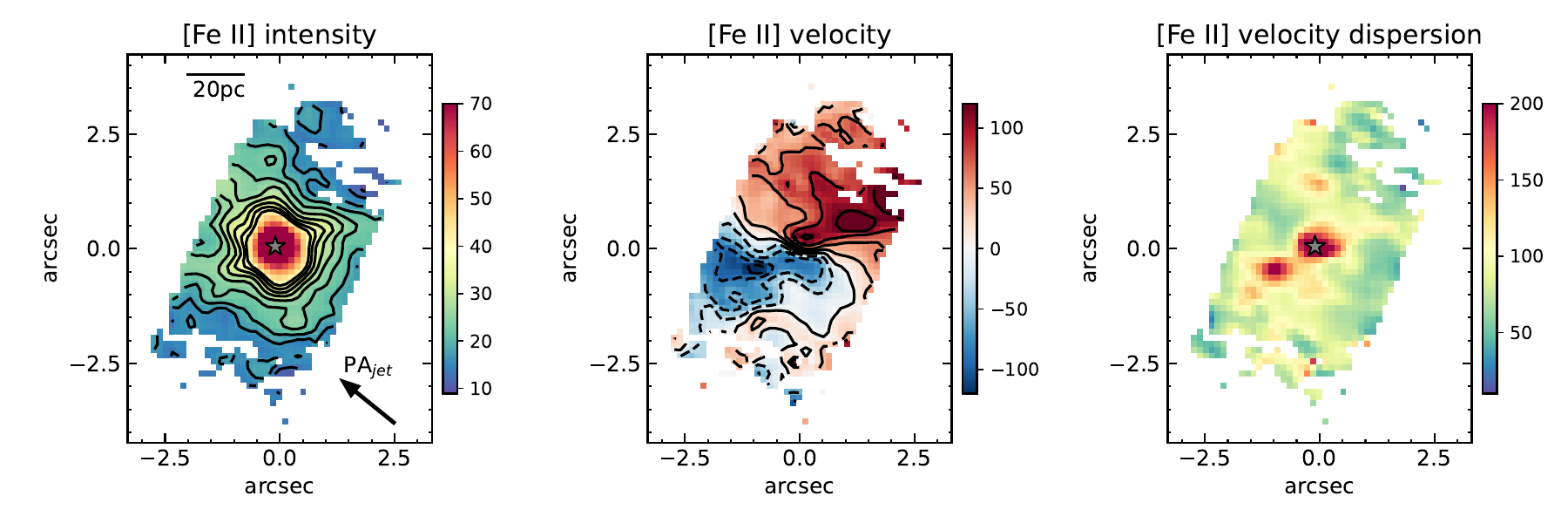}

  \vspace{-0.25cm}
 \hspace{1.5cm} 
  \includegraphics[width=14.7cm]{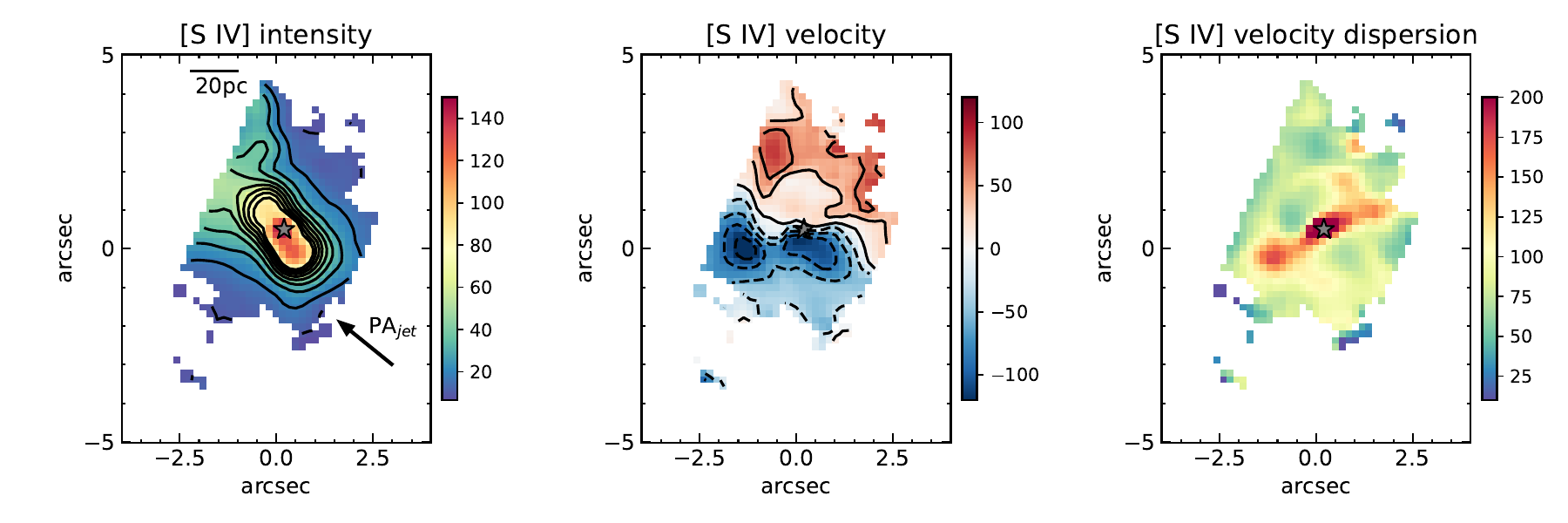}

   \vspace{-0.25cm}
 \hspace{1.5cm} 
  \includegraphics[width=14.7cm]{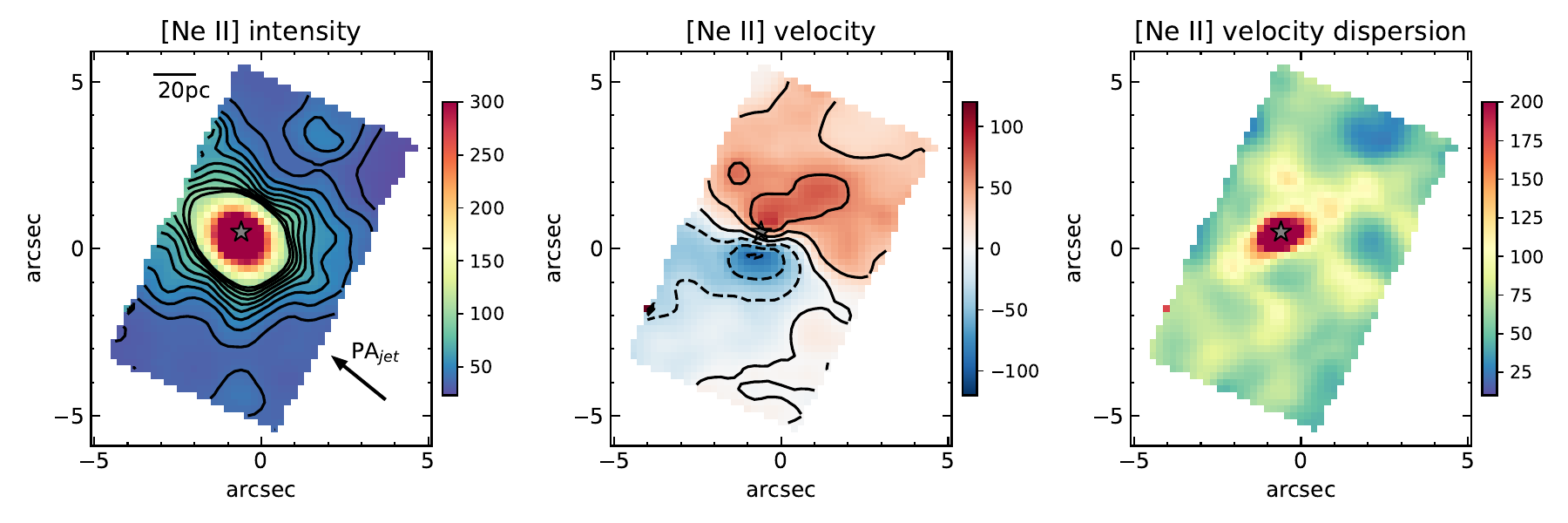}

  \vspace{-0.25cm}
 \hspace{1.5cm} 
  \includegraphics[width=14.7cm]{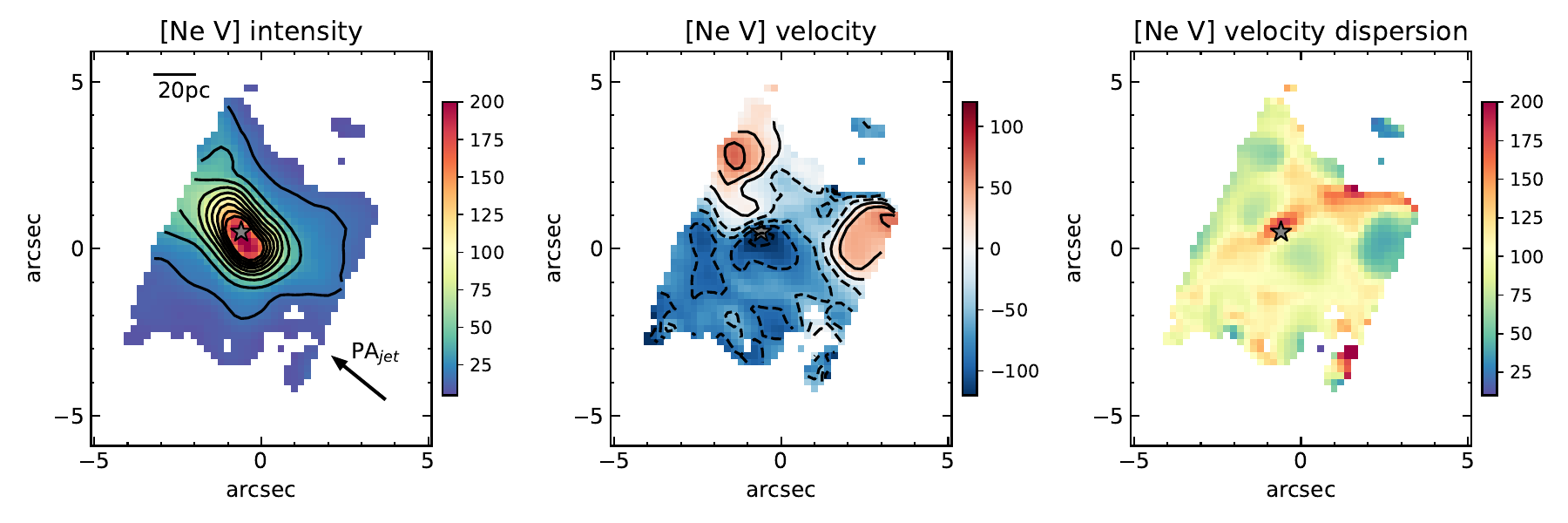}

    \vspace{-0.25cm}
 \hspace{1.5cm} 
  \includegraphics[width=14.7cm]{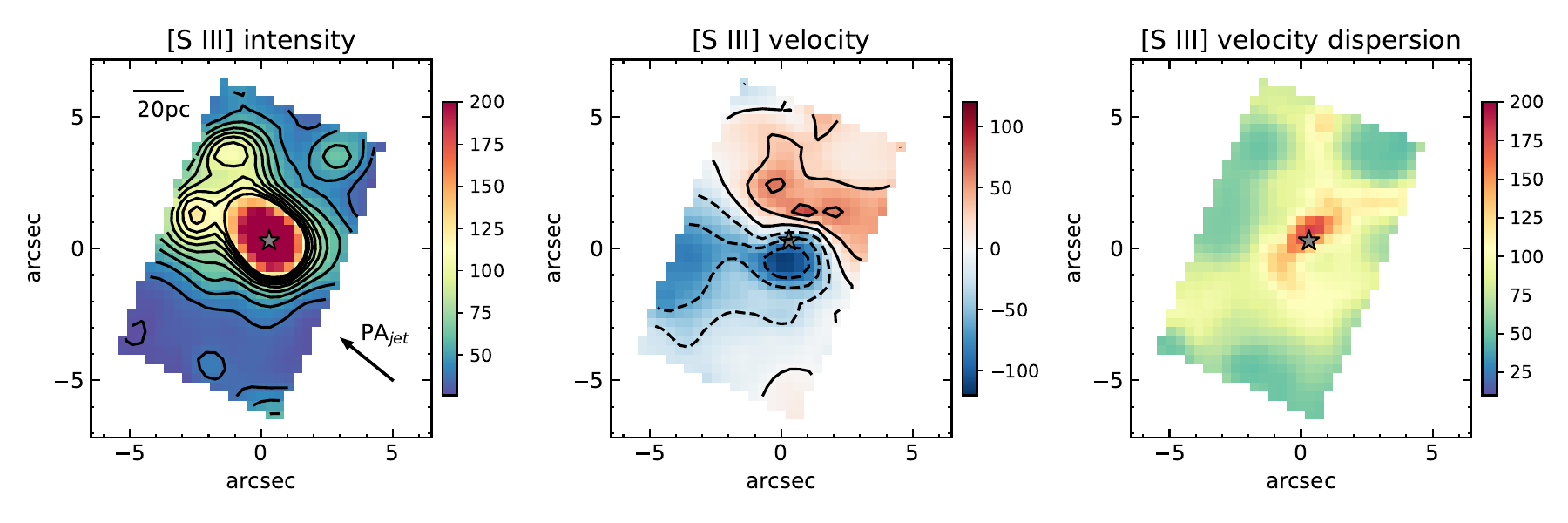}

 \vspace{-0.3cm}
  \caption{Same as Fig.~\ref{fig:maintextalucinemapsFSL} but for  [Fe\,{\sc
      ii}] at $5.34\,\mu$m, [S\,{\sc iv}],  [Ne\,{\sc ii}], [Ne\,{\sc v}] at $14.32\,\mu$m, and
[S\,{\sc iii}]
(from top to bottom).}
     \label{fig:appendixalucinemaps}
\end{figure*}

\begin{figure*}[hbt!]

  \vspace{-0.35cm}
 \hspace{1.5cm} 
 \includegraphics[width=14.7cm]{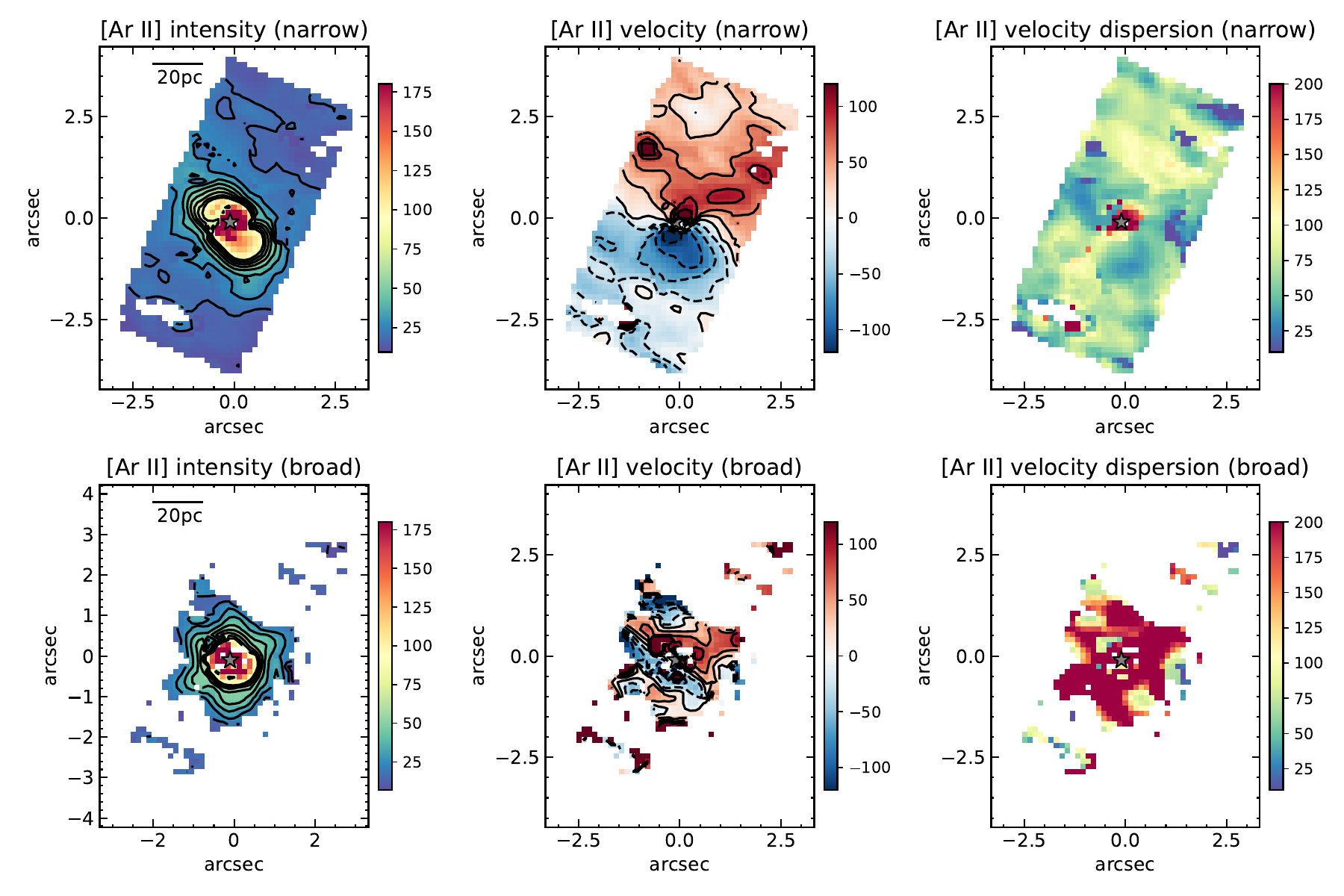}

 \vspace{-0.3cm}
  \caption{Same as Fig.~\ref{fig:maintextalucinemapsFSL} but for fits with two
    Gaussian components for the [Ar\,{\sc ii}] line. All velocities
    are referred to $v_{\rm sys}$.}
     \label{fig:appendixalucinemapstwocomponents}
\end{figure*}

We also show additional maps for lines fit with one
Gaussian component, namely, [Fe\,{\sc ii}] at $5.34\,\mu$m, [S\,{\sc iv}], [Ne\,{\sc ii}],
[Ne\,{\sc v}] at $14.32\,\mu$m, and
[S\,{\sc iii}] in Fig.~\ref{fig:appendixalucinemaps} as well as maps fit with
two Gaussian components for [Ar\,{\sc ii}] in
Fig.~\ref{fig:appendixalucinemapstwocomponents}.  

\begin{figure*}[h!]
  \hspace{3cm}
  \includegraphics[width=10cm]{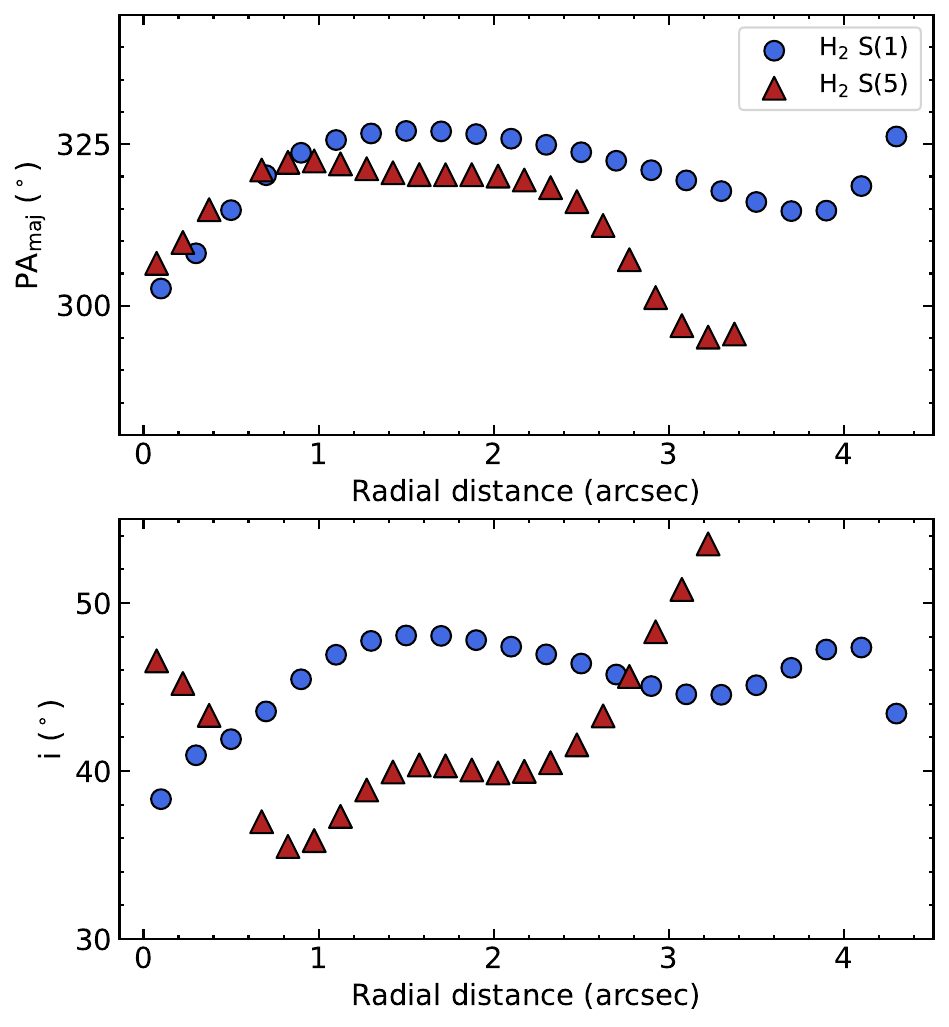}
\caption{Radial dependence of the kinematic PA$_{\rm maj}$  (top) and
  $i$ (bottom) of the  $^{\rm 3D}$BAROLO disk 
  models (no $v_{\rm RAD}$ component) of the H$_2$ S(5) and H$_2$ S(1)
  transitions (top-left panels of
  Fig.~\ref{fig:BAROLOvfields_H2S5} and
  Fig.~\ref{fig:BAROLOvfields_H2S1}, respectively). An average
value of $i \simeq 45^{\rm o}$ was derived for the nuclear disk
and nuclear ring by \cite{Neumayer2007}, in good agreement with our fits.} 
     \label{fig:BAROLOparameters_H2S5NOVRAD}
\end{figure*}

\begin{figure}[h!]
  \hspace{1cm}
  \includegraphics[width=7cm]{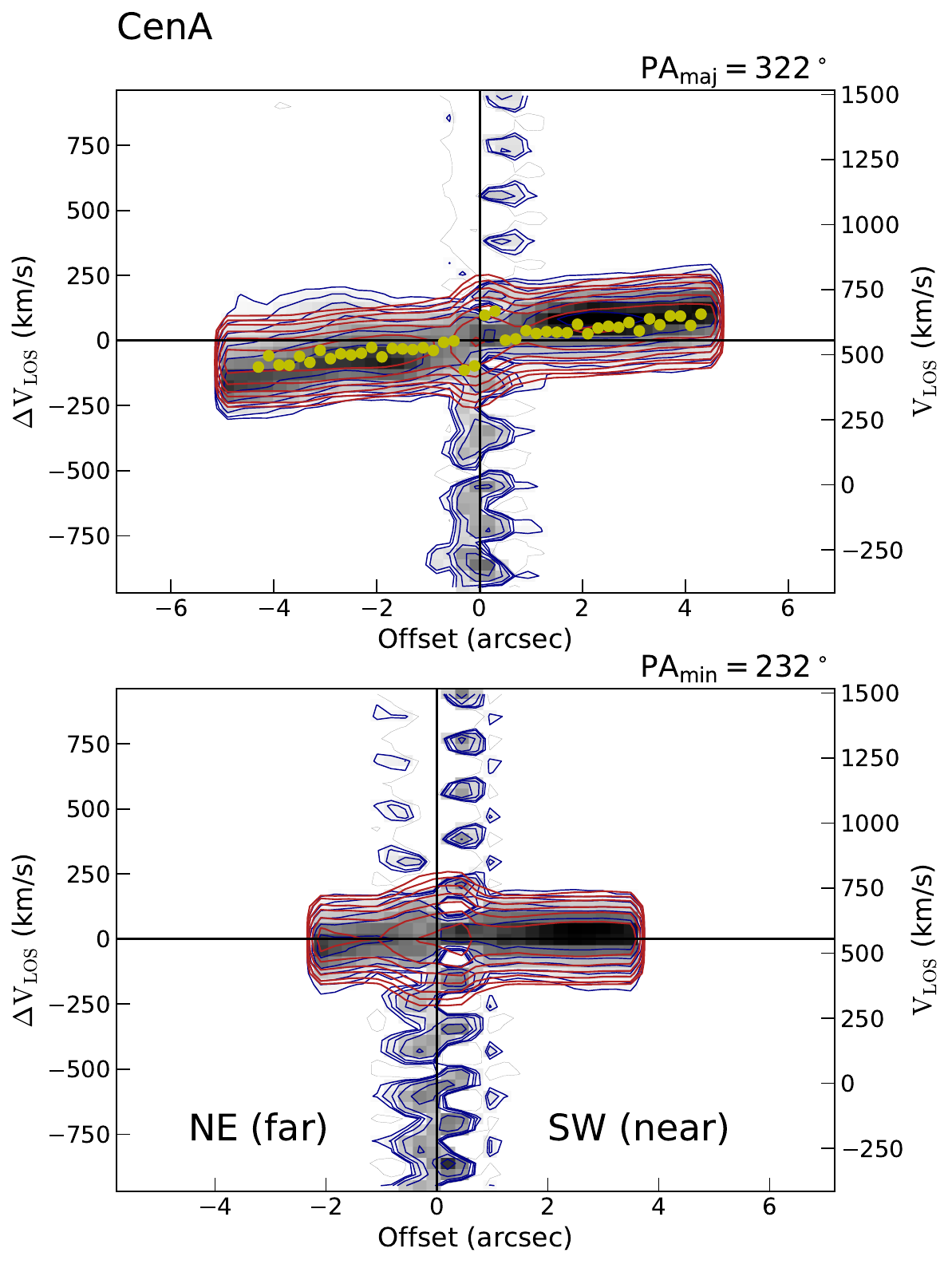}
  \caption{P-v diagrams for H$_2$ S(1) where the disk model has no
    $v_{\rm RAD}$ component. Colors and lines are as in
    Fig.~\ref{fig:BAROLOpv_H2S5NOVRAD}. We call attention to the strong residuals 
left from subtracting the strong continuum at the AGN
       position.} 
     \label{fig:BAROLOpv_H2S1NOVRAD}
\end{figure}

\begin{figure}[hbt!]
\hspace{0.25cm}
  \includegraphics[width=8.5cm]{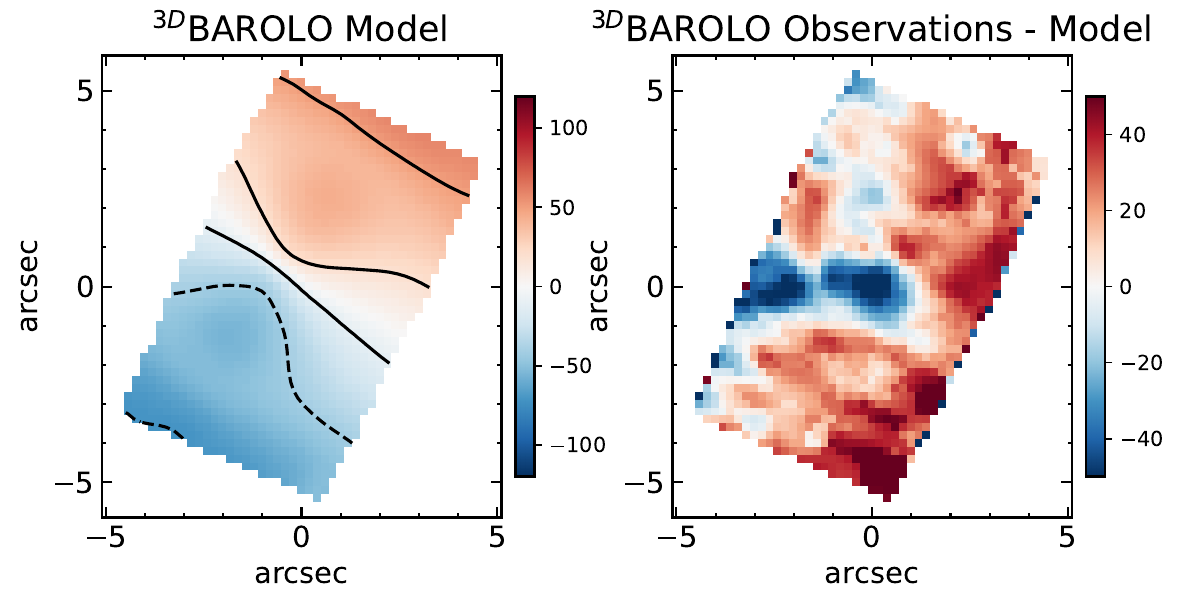}
  \caption{Same as top panel of Fig.~\ref{fig:BAROLOvfields_H2S5}
but for the velocity fields of [Ne\,{\sc iii}].}
     \label{fig:BAROLOvfields_NeIII}
\end{figure}

\begin{figure}[hbt!]
  \hspace{1cm}
  \includegraphics[width=7cm]{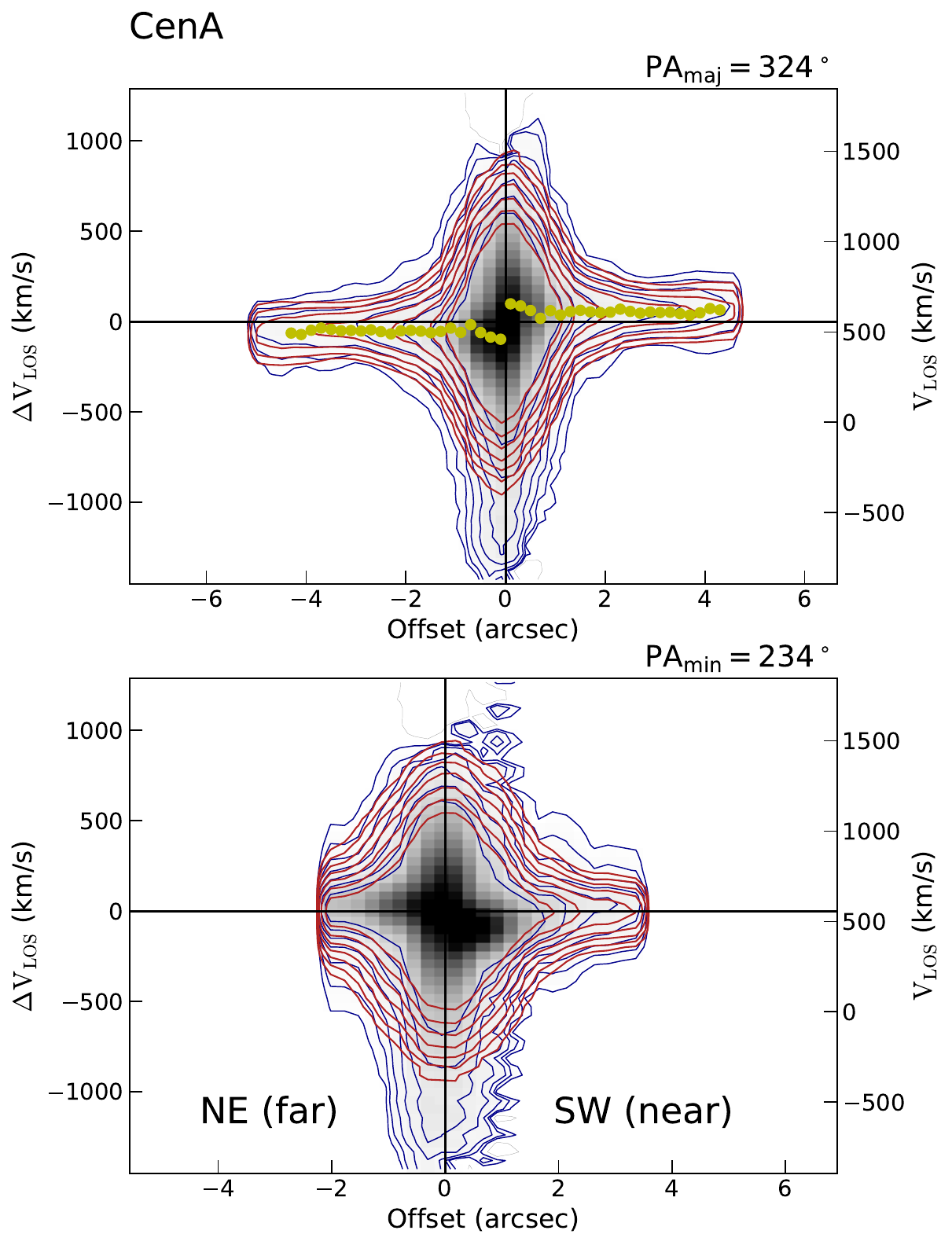}
\caption{P-v diagrams extracted along the kinematic major axis (top)
  and minor axis (bottom) for [Ne\,{\sc iii}]. Colors and 
  contours as in Fig.~\ref{fig:BAROLOpv_H2S5NOVRAD}.}
     \label{fig:BAROLOpv_NeIIINOVRAD}
\end{figure}

Finally, we include additional figures generated with $^{\rm
  3D}$BAROLO. Fig.~\ref{fig:BAROLOparameters_H2S5NOVRAD} compares the
derived PA$_{\rm maj}$ and $i$ of the disk
  models with no $v_{\rm RAD}$ component  of the H$_2$ S(5) and
  H$_2$ S(1) lines, while Figs.~\ref{fig:BAROLOpv_H2S1NOVRAD}, and \ref{fig:BAROLOpv_NeIIINOVRAD}
  are p-v diagrams.  Figure~\ref{fig:BAROLOvfields_NeIII} shows the model and
  residuals for the fit to the [Ne\,{\sc iii}] line.

  \begin{table*}[h!]
    \caption{Parametric fits to the line profiles with one, two, or three Gaussian components.}
  \begin{tabular}{l|ccc|ccccc|ccccccc}
    \hline
    Line & \multicolumn{3}{c}{Single Gaussian} &
                                                 \multicolumn{5}{c}{Two
                                                 Gaussians } &
    \multicolumn{7}{c}{Three Gaussians}\\
    & $v_1$ & $\sigma_1$ & $\epsilon$ & $v_1$ & $\sigma_1$ &
                                                                  $v_2$
    & $\sigma_2$ & $\epsilon$ &
    $v_1$ & $\sigma_1$ & $v_2$ & $\sigma_2$& $v_3$ & $\sigma_3$ & $\epsilon$ \\
            & \multicolumn{2}{c}{(km s$^{-1}$)} & &
                                                   \multicolumn{4}{c}{(km s$^{-1}$)} & & \multicolumn{6}{c}{(km s$^{-1}$)} \\
    \hline
    H$_2$ S(5) & 23.1 & 83.2& 1.6 & 42.5 & 58.1 & -1.2 & 101.6 & 1.1\\
    ${\rm [Ar\,II]}$ & 1.6 & 258.2 & 5.7 &
                              -7.2 & 188.5 & 36.3 & 567.4 & 3.4 &
  -71.3 & 41.7 &  26.9 & 236.7 & -55.3 & 701.5  & 1.0\\
    ${\rm[Ne\,III]}$ & -11.8 & 167.5 & 37.5 &
                               -25.0 & 116.0 & 39.2 & 306.6 & 11.6 &
-32.8 & 110.0 & 47.4 & 231.0 & -112.7 & 605.6 & 5.8\\
    \hline
  \end{tabular}
  Notes.--- The line profiles are those from the spectrum
    extracted as a point source
    (Sect.~\ref{subsec:analysis}). For each component $v$ is the peak velocity and $\sigma$
  the velocity dispersion. All the velocity dispersions are corrected for
  instrumental resolution. $\epsilon$ is an estimate of the fit
  residual (see Section~\ref{subsec:analysis} for details).
  \label{tab:parametricfits}
  \end{table*}

\FloatBarrier 
\twocolumn

\FloatBarrier 

\end{appendix}

  \end{document}